\documentclass[12pt,a4paper]{article}

\usepackage{eurosym}
\usepackage{amscd,amsfonts,amsmath,amssymb,amsthm,bbm,bm,latexsym,mathrsfs}
\usepackage{epsfig,graphics,graphicx,natbib,subfigure,longtable}
\usepackage[english]{babel}
\usepackage[usenames]{color,colortbl}

\usepackage{bookmark}
\usepackage{booktabs}
\usepackage{rotating}
\usepackage{wasysym}
\usepackage{lscape}
\usepackage[flushleft]{threeparttable}
\usepackage{adjustbox}

\usepackage{sgame}
\usepackage[normalem]{ulem}% to highlight cancelled text
\usepackage{lineno} % for line numbering
\usepackage[top=1in, bottom=1in, left=0.5in, right=0.5in]{geometry}
\usepackage{setspace}
\makeatother

\RequirePackage{xr}
\theoremstyle{remark}

\theoremstyle{plain}

\usepackage{ragged2e}

\begin{document}
%\linenumbers

%\newgeometry{top=1.1in, bottom=1.1in, left=1.2in, right=1.2in}
\newgeometry{top=1in, bottom=1in, left=0.7in, right=0.7in}

\title{\vspace{-50pt} Comparing the Forecasting Performances of Linear Models for Electricity Prices with High RES Penetration}

\author{
\hspace{-20pt}
Angelica Gianfreda\textsuperscript{a,c} \hspace{15pt}
Francesco Ravazzolo\textsuperscript{a,d} \hspace{15pt}
Luca Rossini\textsuperscript{b,a}
 \\
 \\
        {\centering {\small
        \textsuperscript{a}Free University of Bozen-Bolzano, Italy \hspace{5pt} \textsuperscript{b}Vrije Universiteit Amsterdam, The Netherlands}} \vspace{5pt} \\
      {\centering {\small \textsuperscript{c}EMG, London Business School, UK \hspace{5pt} \textsuperscript{d}CAMP, BI Norwegian Business School, Norway}}
     }

\date{\today}%\date{}
\maketitle

\doublespacing

%%%%%%%%%%%%%%%%%%%%%%%%%%%%%%%%%%%%%%%%%%%%%%%%%%%%%%%
%% SUMMARY REQUIRED
%%%%%%%%%%%%%%%%%%%%%%%%%%%%%%%%%%%%%%%%%%%%%%%%%%%%%%%
%
%Summary: Please provide a summary of no more than 100 words. A summary is a condensation of the whole paper, not just the conclusions, and is understandable without reference to the rest of the paper. It should contain no citation to other published work.
%

%%%%%%%%%%%%%%%%%%%%%%%%%%%%%%%%%%%%%%%%%%%%%%%%%%%%%%
%%         		  	Abstract		   				  %%%
%%%%%%%%%%%%%%%%%%%%%%%%%%%%%%%%%%%%%%%%%%%%%%%%%%%%%%
\abstract{
This paper compares alternative univariate versus multivariate models, frequentist versus Bayesian autoregressive and vector autoregressive specifications, for hourly day-ahead electricity prices, both with and without renewable energy sources. The accuracy of point and density forecasts are inspected in four main European markets (Germany, Denmark, Italy and Spain) characterized by different levels of renewable energy power generation. Our results show that the Bayesian VAR specifications with exogenous variables dominate other multivariate and univariate specifications, in terms of both point and density forecasting.\\


%\noindent \textbf{JEL codes:} C11, C53, C55, Q42, Q47.
}

%%%%%%%%%%%%%%%%%%%%%%%%%%%%%%%%%%%%%%%%%%%%%%%%%%%%%%
%%         		Introduction		   				  %%%
%%%%%%%%%%%%%%%%%%%%%%%%%%%%%%%%%%%%%%%%%%%%%%%%%%%%%%
\section{Introduction}
\label{sec_Intro}

Despite the recent availability of high frequency data for forecasted demand and renewable generation, the literature on forecasting electricity prices using these exogenous variables is still relatively scarce. Therefore, we aim to fill this gap by looking at linear models, in both univariate and multivariate frameworks, while comparing the frequentist with the Bayesian approach and evaluating both point and density forecasts.

This paper shows that hourly prices can be predicted efficiently by taking advantage of intra-daily information available to market participants when controlling for fossil fuels. We have explored linear autoregressive (AR) and vector autoregressive (VAR) models, both with and without fundamental predicted drivers (forecasted demand, forecasted wind and solar power generation). These exogenous variables play an important role in formulating day-ahead conditional expectations, and their effects have motivated extensive research. Furthermore, in the last ten years, electricity generated from renewable energy sources (RES-E) has grown significantly thanks to the political and financial support for these sources, which may play an essential role not only in reducing country energy dependence (on imported fossil fuels) but also, and more importantly, in mitigating global warming (by reducing greenhouse gas emissions). The renewable energy sources' (RES) share of the total power capacity increased from 24\% to 44\% between 2000 and 2015 in Europe, reaching a total of more than 2,000 GW in 2016. The share of wind power increased from 2.4\% to 15.6\%, with a total generation of approximately 300 TWh, covering more than 10\% of EU demand. Denmark and Germany were among the leading countries for total wind power capacity per inhabitant. %Notwithstanding ongoing economic crises, austerity measures and retroactive policy changes, Spain ranked second in Europe for total wind operating capacity (with more than 20GW in 2015); whereas, Italy contracted significantly its wind annual installations in 2016. However, high RES penetration levels were observed: 38\% in Denmark, 19\% in Spain, and 13\% in Germany (with 11\% from onshore and 2\% from offshore). Instead lower shares were observed for solar PV which accounted for more than 7\% in Italy, and 6.4\% in Germany.
The global solar PV capacity totalled an estimated 106 GW in Europe at the end of 2016, which is more than 32 times the capacity observed in 2006. Germany, Italy, and Spain are found to belong to the group of top ten world countries for capacity and additions (see \citealp{REN21}). These statistics support our choice of selected markets. %; with Germany and Italy being also the leaders for solar PV capacity per inhabitant

On the operational side, RES have added complexity to the management of the electricity system, and, thus to electricity price modelling and forecasting. % because of their high variability and partial predictability.
 Consequently, a growing body of literature has investigated the effects of RES on electricity price dynamics in several markets around the world (Europe, United States, Canada, and Australia). Given the uncertainties in the forecasted levels of demand and RES-E, market operators are concerned about the forecasts of day-ahead prices.

Still, there is no empirical consensus about the superiority of multivariate versus univariate models, and we aim at filling this gap when all fundamental drivers are considered, thus providing clear operational guidelines in forecasting hourly day-ahead electricity prices. Therefore, this paper compares various univariate and multivariate linear models with and without RES-E forecasts and other fundamental drivers, estimated using frequentist and Bayesian approaches, for producing day-ahead forecasts of selected European electricity prices. Indeed, the advent of RES has raised numerous challenges for electricity markets in terms of managing, monitoring, modelling and forecasting. Renewables (as wind and solar) have zero marginal production cost but are intermittent: if the wind blows and/or the sun shines, electricity prices are low; otherwise, when the sun stops shining or the wind stops blowing, traditional thermal plants running with fossil fuels must produce demanded electricity with higher generation costs. Consequently, some negative prices can arise when power from RES is sufficient to meet demand and some units must be paid to reduce production and/or increase demand.\footnote{Negative prices are considered market signals of inflexibility: the system is not able to increase the demand on one hand and to reduce generation on the other hand, because turning conventional power plants on and off would be inefficient and uneconomical.} Therefore, this emphasizes the importance of including RES-E and other fossil fuels when looking for the best price forecasts. While there is unanimous consensus that including demand forecasts or RES-E forecasts (if the market penetration is not negligible) leads to more accurate forecasts, it is still an open question as to which RES-E forecast is more informative in which market, and also whether their inclusion can reduce the importance of fossil fuels. Hence, models with only a subset of exogenous variables have been considered also.

Our results show that demand and renewable energies improve the point and density accuracy of the predictive models, especially during peak hours. However, their inclusion does not reduce the importance of fossil fuels, which we suggest should be retained in the models. Moreover, we find evidence of better forecasting of the multivariate models, given that they allow for interrelationships among different hours of the day, and the Bayesian approach leads to further forecasting improvements. Finally, and for the first time since the increasing RES penetration, we show that the models with forecasted wind only (besides forecasted demand and fuels) perform better than do those with solar power only (besides forecasted demand and fuels). And, their simultaneous inclusion further improves the performance.

The paper proceeds as follows. Section \ref{sec_LitRev} summarizes previous research on forecasting electricity prices and highlights our contributions. Section \ref{sec_Data} contains the description of the market together with details on the data used. Section \ref{sec_Models} presents our models, estimation methodology, and the metrics used to assess our results. These are discussed in Section \ref{sec_Results}, together with the major findings. Finally, Section \ref{sec_Conclusions} concludes.

\section{Literature Review}
\label{sec_LitRev}

As emphasized in two reviews by \cite{Weron2014} and \cite{Nowotarski2018}, there is increasing interest in electricity price forecasting. However, few studies have addressed the comparison of univariate and multivariate models within the frequentist and Bayesian approaches, when considering both point and density forecasts and the forecasting ability of fundamental drivers. In performing extensive empirical comparisons, we aim to fill this gap while exploring several combinations between forecasted variables and fossil fuel prices.

Several studies have considered the univariate dimension for modelling purposes, e.g. \cite{Koopetal2007}; \cite{Karakatsani2008}; \cite{Gianfreda2012}; and \cite{Chen2014}. However, they did not include any forecasted renewable power generation. And, more recent papers have analysed the impact of RES on wholesale electricity price dynamics, see \cite{Jonsson2010}, \cite{Gelabert2011}, \cite{Wooetal2011}, \cite{Mauritzen2013}, \cite{Ketterer2014}, \cite{Paraschiv2014}, \cite{Brancucci2016},  \cite{Pircalabu2017} and \cite{Rintamaki2017}, among many others. It is worth emphasizing that most of the authors have modelled each hourly time series individually (that is 24 hourly time series separately), as in \cite{Misiorek2006} and in \cite{GarciaMartos2007}, hence, ignoring the relationships among different hours of the day.

To overcome this issue, \cite{Maciejowska2016a} proposed 24 separate autoregressive models, including, among the regressors of the models, the early morning hours (up to 4 a.m.), the last prices (at hours 23 and 24) from the previous day, historical prices (at lags 1 and 7), a weekend dummy to capture seasonality, and load selected again at lags 1 and 7; however no RES were included. In addition to AR models, \cite{Maciejowska2015} also proposed VAR models for hourly and averaged daily prices, with 480 estimated parameters for working/weekend days, daylight hours, and a constant 7-lag order structure;  which still does not involve demand and renewable power.

Therefore, following \cite{Conejo2005}, \cite{Misiorek2006}, and \cite{Maciejowska2015}, we select AR models as benchmarks because of their widespread use in the literature and their relatively good performance in predicting electricity prices. Moreover, we consider VAR representations to detect improvements in the forecasting performances. Indeed, we expect better forecasts from multivariate than univariate models given the larger information contained in a panel of data, as suggested by \cite{Stock2002}.

Being aware of the explosion in dimensionality, we push these models forward by including also forecasted demand and RES-E in both our univariate AR and multivariate VAR models. Furthermore, we consider exploring natural gas, coal, and CO$_2$ if their inclusion improves the forecasting ability, as shown by \cite{MacWer2016}. Hence, we manage a total of 161 parameters for each hour.

As far as forecasting is concerned, and has emerged from the reviews, few studies have considered density forecasting (e.g. \cite{Panagiotelis2008}; \cite{Huurman2012}; \cite{Jonsson2014}; and \cite{Gianfreda2018}). %and \cite{Ziel2016}.

More recently, but without accounting for fundamental drivers and looking only at point forecasts, \cite{Ravivetal2015} compared the performances of models for the full panel of 24 hourly prices studying NordPool from 1992 to 2010. Based on univariate AR and multivariate VAR models, they computed forecast combinations and empirically demonstrated that the useful predictive information contained in disaggregated hourly prices improves the forecasts of multivariate models. They showed that shrinking VAR models leads to further better forecasts, with the Bayesian VAR outperforming the unrestricted VAR. However, no density forecasting was performed and no RES were included in their models, as in \cite{ZielWeron2018}.  \cite{ZielWeron2018} proposed 58 multi-parameter regression univariate and multivariate models accounting for different forms of seasonality, but no evidence of the uniform superiority of multivariate specifications was provided across all 12 studied markets, seasons or hours. More specifically, and closer to our analysis, they concluded that, in Spain, the multivariate specification often outperforms the univariate specification in the morning hours, whereas, in Germany and in the two Danish zones, the univariate specification often outperforms the multivariate specification in the late evening/night hours. However, these results depend on the specifications of their models and may produce different results if forecasted demand and RES-E are included. Therefore, this further supports our investigation and we aim at providing even more clear evidence on linear univariate and multivariate forecasting performances comparing frequentist and Bayesian models when more complexity is induced by uncertain and intermittent renewable generation.
%%
%%%%%%%%%%%%%%%%%%%%%%%%%%%%%%%%%%%%%%%%%%%%%%%%%%%%%%%
%%%         		  		Data		   				  %%%
%%%%%%%%%%%%%%%%%%%%%%%%%%%%%%%%%%%%%%%%%%%%%%%%%%%%%%%
%
\section{Market Structure and Data Description}
\label{sec_Data}

\subsection{The Electricity Market and its Sessions}

Wholesale electricity markets are platforms where electricity is traded. These are organized in sequential sessions: the day-ahead, intra-day, and balancing sessions. In the day-ahead session, bids to buy and offers to sell electricity for each hour of the following day are submitted in pairs of prices and quantities by consumption units and generators on a voluntary basis (there is no obligation to act).
This session opens several days in advance and closes one day before physical delivery. For this reason, these markets are often called forward, auction, or day-ahead markets, in which individual supply offers and demand bids are ordered giving priority of dispatch to more efficient and less polluting units with lower marginal costs (then wind and solar -- RES in general -- enter the supply curve before nuclear, coal, and gas units, which have higher marginal costs; this is the so called `merit order criterion').
Hence, the price is computed under a cost minimising objective on an hourly basis and it is identified by the intersection of the aggregated curves of supply and demand. This day-ahead price is determined according to generators' planned schedules of production and by forecasted consumption programmes, which can be affected by sudden outages and weather conditions among many other factors.

Subsequently, the intra-day sessions take place, wherein units are allowed to modify (by buying or selling) their day-ahead schedules as new information (like better weather forecasts) becomes available. These operations are undertaken generally by units of intermittent and variable generation (but recently also some thermal units have started to play across day-ahead and intra-day sessions to explore higher profit opportunities in balancing sessions where prices are higher and the price-as-bid is used). The participation at the day-ahead and intra-day sessions occurs on a voluntary basis, they are both managed by the system operator and a marginal pricing rule applies.

The balancing sessions represent the last sessions used by the transmission system operator to grant system security and grid stability and to match instantaneously demand and supply in case of any unexpected imbalance. These are usually organized in an `ex-ante' planning phase (when generation resources are committed) and in a `real-time' session (when the balancing is granted to restore frequency and quantity deviations); hence, several types of products are actually remunerated. Given that only generators with the required degree of flexibility are allowed to provide these services, these sessions are generally more concentrated than are the former ones, the participation is mandatory, and the pay-as-bid pricing mechanism is applied (for additional details, see \citealp{Hirth2015} and \citealp{PopVri2019}).

Then, day-ahead forecasts are particularly important for the market itself and for operators, because, if the day-ahead forecasts (of quantities) are wrong, then energy must be acquired in the real-time market at a (potentially and generally) higher price, as highlighted by \cite{Gianfredaetal2018}, who investigated all these market sessions and the bidding behaviour of (hydro, water pumping and thermal conventional) balancing responsible units in the Northern zone of Italy.

Given the uncertainties in the forecasted levels of demand and, more importantly, those in the forecasted levels of RES-E (affecting the supply curve according to the levels of RES penetration), substantial variability is introduced. And this also explains why one step ahead forecasts are gaining increasing interest. Moreover, market operators and traders are concerned about these forecasts of day-ahead prices because they are used in the balancing pricing mechanisms and can provide an indication of the magnitude of price spreads across sessions (see \cite{Bunnetal2018} and \cite{LisiEdoli2018} for further details about imbalances and strategic speculations).

\subsection{Data}

We use hourly day-ahead prices (in levels) to estimate models for electricity traded/sold in Germany, Denmark, Italy, and Spain. These markets are particularly interesting, given their high levels of RES penetration. Following \cite{Uniejewskietal2016} and \cite{ZielWeron2018}, we refer to day-ahead and spot interchangeably to identify prices determined in a market today for delivery in a certain hour tomorrow, according to the literature on European electricity markets. Formally, they are forward prices determined one day in advance and with maturity in the following day.\footnote{However, it must be emphasized that in the US the spot market is used to indicate the real-time market, whereas the day-ahead market is usually and more properly called the forward market.}
This time difference is important in understanding the usage of forecasted variables (as demand, wind, and solar) available to operators when they run their forecasting models to obtain a set of 24 prices to be submitted to power exchanges before the closure of the market.\footnote{Hence, we are not considering real-time prices determined by balancing needs to match instantaneously demand and supply. These prices are usually called `balancing' prices and determined in other market sessions regulated by different pricing mechanisms (for further insights see \citealp{Hirth2015}; \citealp{Gianfredaetal2019} and \citealp{Gianfredaetal2018}).}
We obtained national electricity prices directly from the corresponding power exchanges: the German hourly auction prices of the power spot market from the \emph{European Energy Exchange} EEX\footnote{Precisely, we had access to the ftp from www.eex.com thanks to the \emph{Europe Energy}}; the two-hourly zonal prices for Denmark from \emph{Nordpool}\footnote{https://www.nordpoolgroup.com} (these were averaged to obtain a single price series for the whole country); the Italian hourly single national prices (\emph{prezzo unico nazionale}, PUN) from the Italian system operator, \emph{Gestore dei Mercati Energetici} GME\footnote{http://www.mercatoelettrico.org}; and the \emph{precios horario del mercado spot diario} for Spain from the \emph{Operador del Mercado Ib\'{e}rico, Polo Espa\~{n}ol}, OMIE\footnote{http://www.esios.ree.es}. These hourly electricity prices (quoted in \euro/MWh), with daily frequency, have been pre-processed for time-clock changes to exclude the 25th hour in October and to interpolate the missing 24th hour in March; hence, there are no missing observations.

As main drivers, we considered both supply and demand sides. As far as the supply side is concerned, we downloaded from Datastream and interpolated missing weekends and holidays of daily settlement prices for coal (as for the Intercontinental Exchange API2 cost, insurance and freight Amsterdam, Rotterdam and Antwerp, with ticker LMCYSPT), for carbon emissions (as for the EEX-EU $CO_2$ Emissions E/EUA in \euro, with ticker EEXEUAS), and for natural gas prices (as for the ICE UK, as it represents a pure hub benchmark and can be used for all EU markets, as suggested by \citealp{Giaetal2016a}) all converted in \euro/MWh using the USEURSP rates from US\$ to Euros (WMR\&DS).
In addition, we consider the forecasted renewable generation (from wind and solar photovoltaic). We downloaded forecasted values for RES-E and demand directly from the market transmission system operators, apart for the German and Italian forecasts, which were provided by Thomson Reuters at hourly frequency.
In these two latter cases, the results from two weather providers (the \emph{European Centre for Medium-Range Weather Forecast} - EC or ECMWF - and the \emph{Global Forecast System} - GFS - of the American weather service of the National Centers for Environmental Prediction) have been inspected.\footnote{Both use two types of weather models: the \emph{operational} one, which is deterministic, with no involved randomness and high resolution; and the \emph{ensemble} one, which is a probabilistic model, with lower resolution and variations around the initial set of weather conditions, hence providing different weather scenarios and, consequently, an idea of the weather instability. Both providers use one single run for the operational model and different runs for the ensemble at specific hours.} % 51 runs (for EC) and 21 runs (for GFS) for the ensemble, respectively. EC runs both (the operational and ensemble) models at 00 and 12, whereas GFS runs them at 00, 06, 12, 18.
We decided to use only forecasts obtained with the EC operational model running at midnight, because this model updates from 05.40 a.m. to 06.55 a.m., thus representing the latest information available to market operators to formulate their day-ahead bidding strategy.

While demand forecast models make use of weather forecasts accounting for temperature, precipitation, pressure, wind speeds, and cloud cover or radiation, forecasted wind values are obtained using the information on wind speeds and installed capacity. Finally, forecast solar power production only considers PV installations, solar radiation, and installed capacity, given the predominance of photovoltaic plants over solar thermal ones. It is worth recalling that the time series for solar power exhibits a block structure of null values in hours early in the mornings and late in the evenings, creating collinearity issues. Hence, we pre-processed these series by a linear transformation: drawing from a Uniform distribution and adding these small numbers to the original zero values in the series. This results in having (column) blocks of very small values close to but different from zero, instead of having (column) blocks of zeros.
%Hence we pre-processed these series by a linear transformation and draw from an Uniform, $\mathcal{U}(0+\delta,0+\varepsilon)$, where $(\delta, \varepsilon)>0$ are small numbers in the neighbourhood of zero. Instead of having columns of zeros, we have columns of different small values close to zero.

\begin{figure}[h!]
\centering
\begin{tabular}{cc}
\includegraphics[width=6cm]{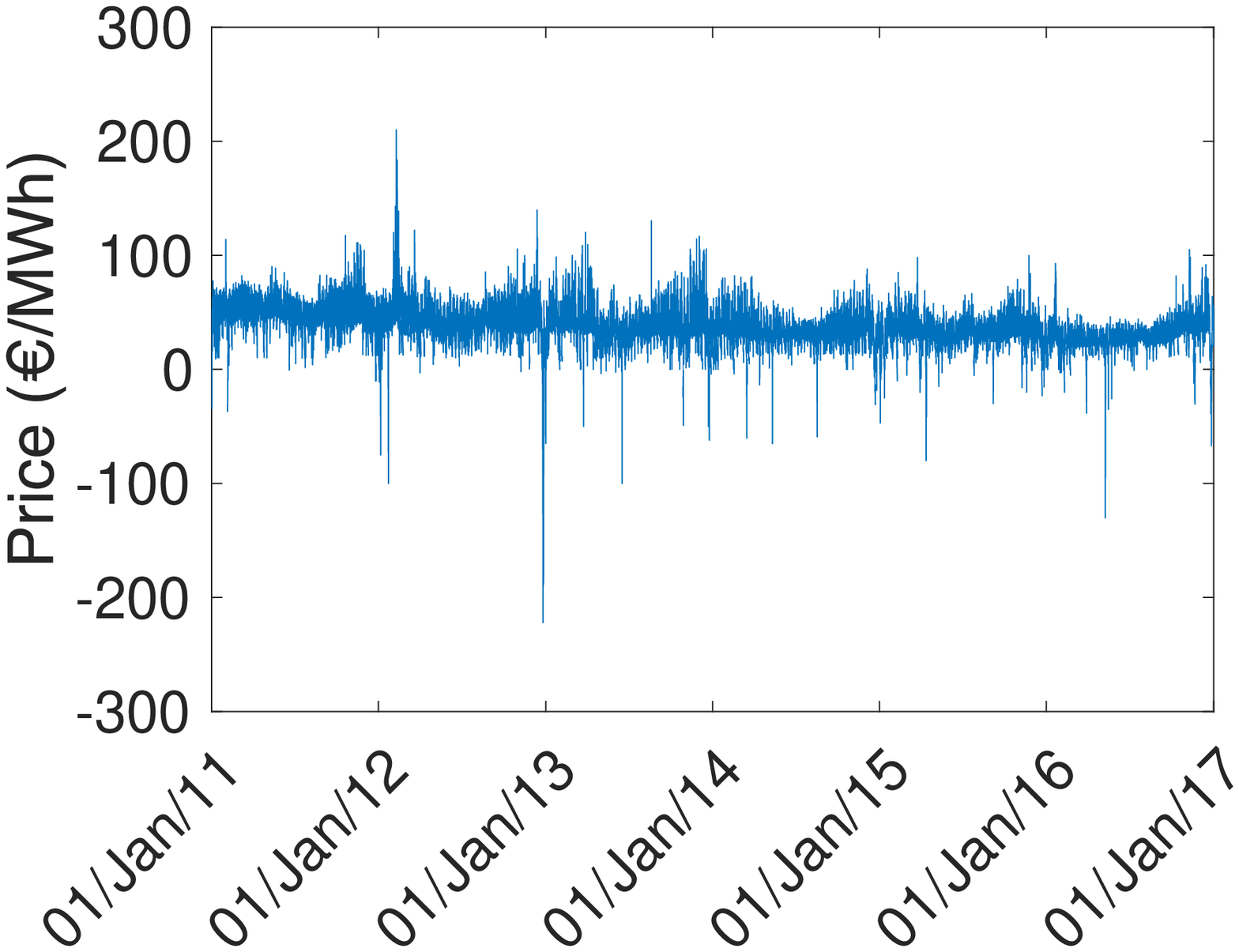}&
\includegraphics[width=6cm]{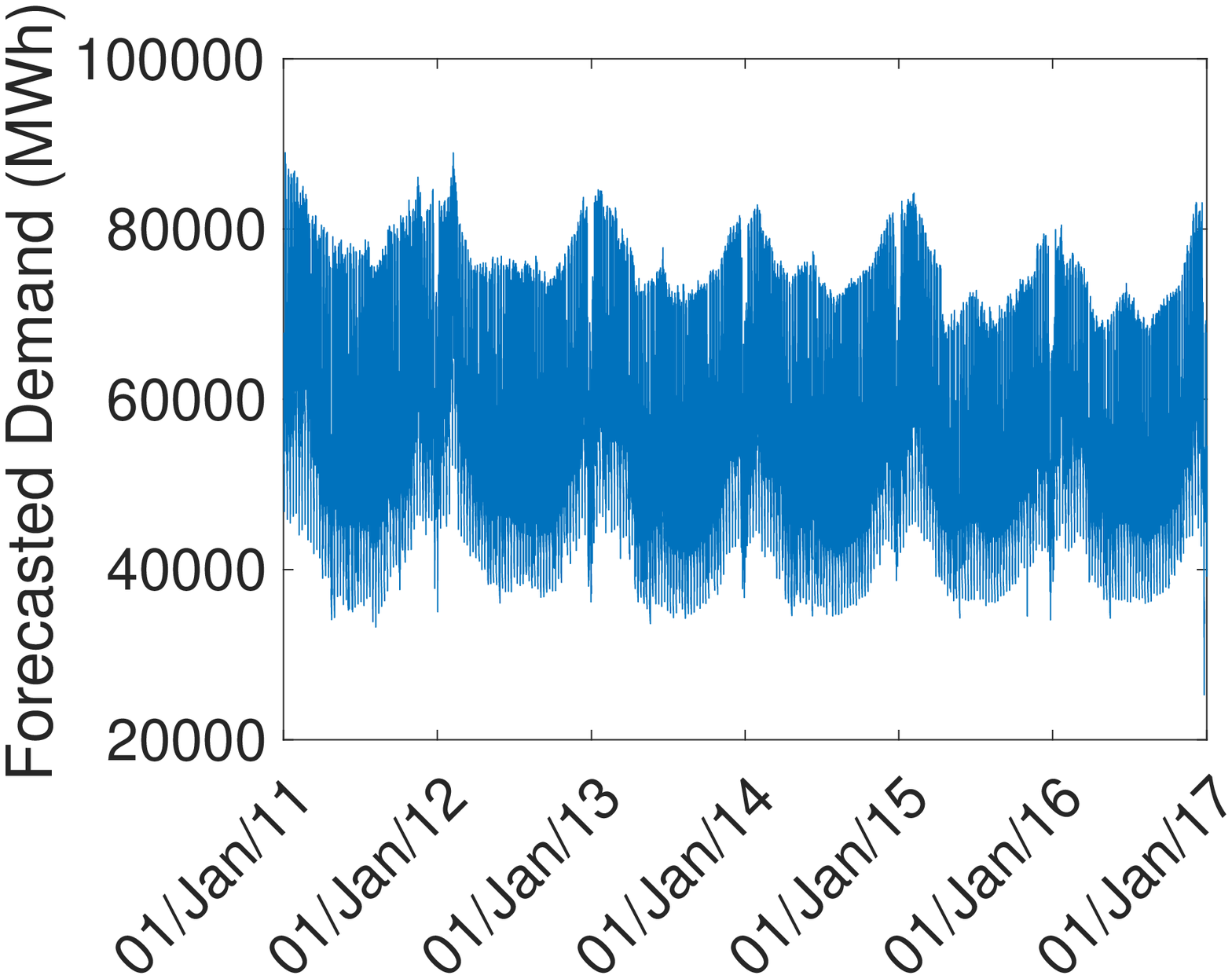}\\
%(a)&(b)\\
\includegraphics[width=6cm]{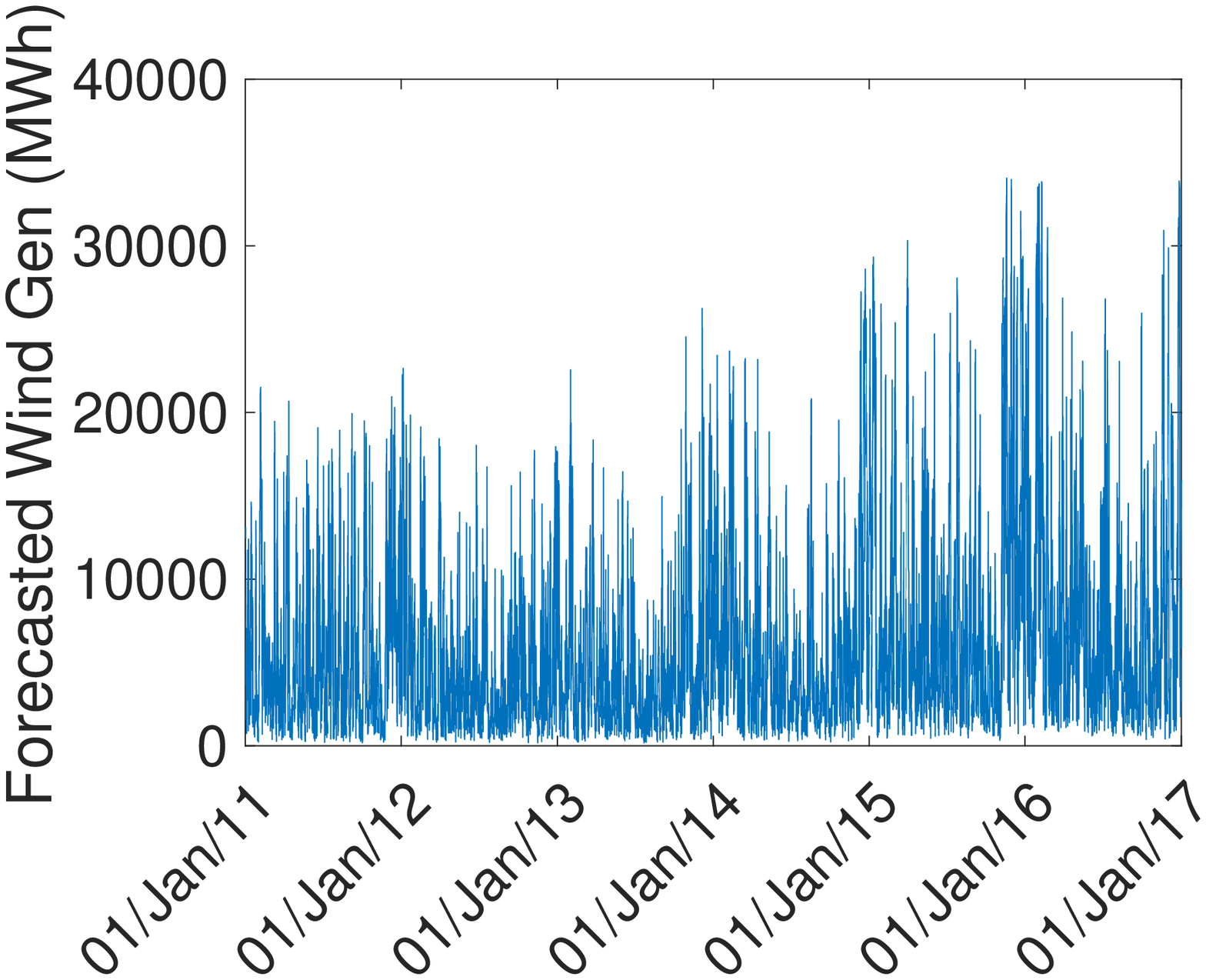}&
\includegraphics[width=6cm]{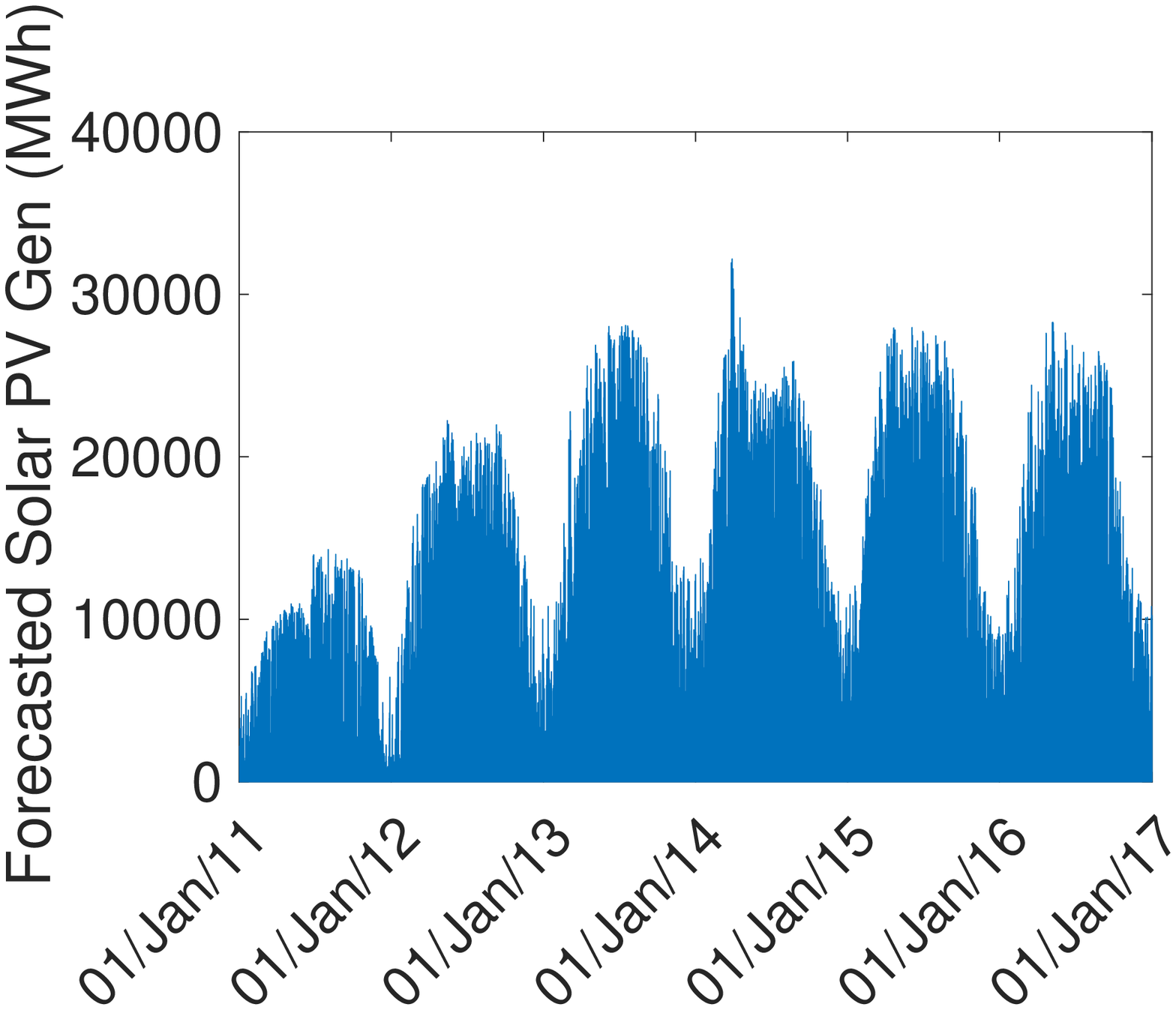}\\
%(c) & (d) \\
\end{tabular}
\caption{\small{Hourly Series for Electricity Day-ahead Prices (top left), Forecasted Demand (top right), Forecasted Wind Generation (bottom left), and Forecasted Solar PV Generation (bottom right) observed in Germany from 01/01/2011 to 31/12/2016.}}
\label{Plot_Germany}
\end{figure}

\begin{figure}[h!]
\centering
\begin{tabular}{cc}
\includegraphics[width=5cm]{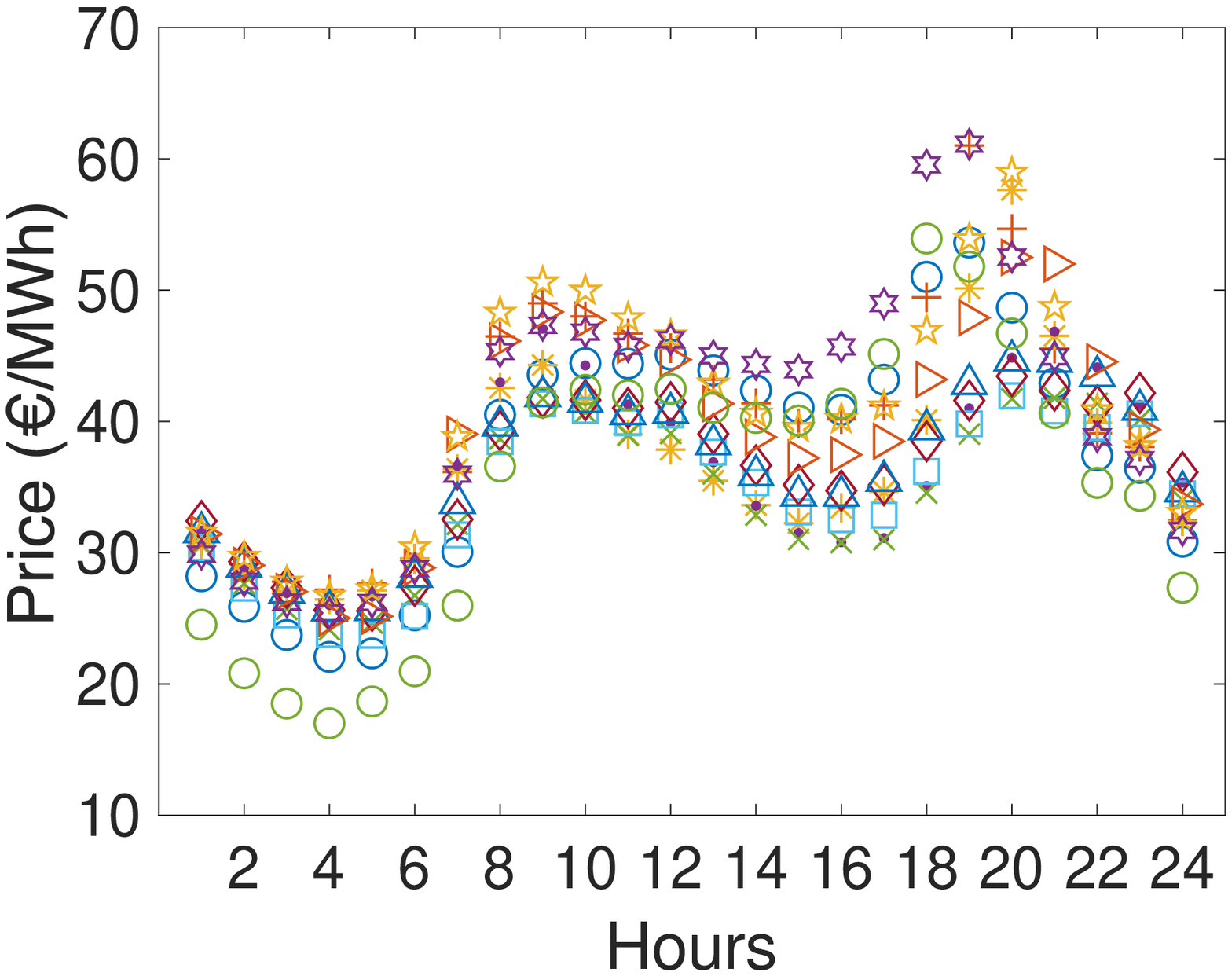}&
\includegraphics[width=5cm]{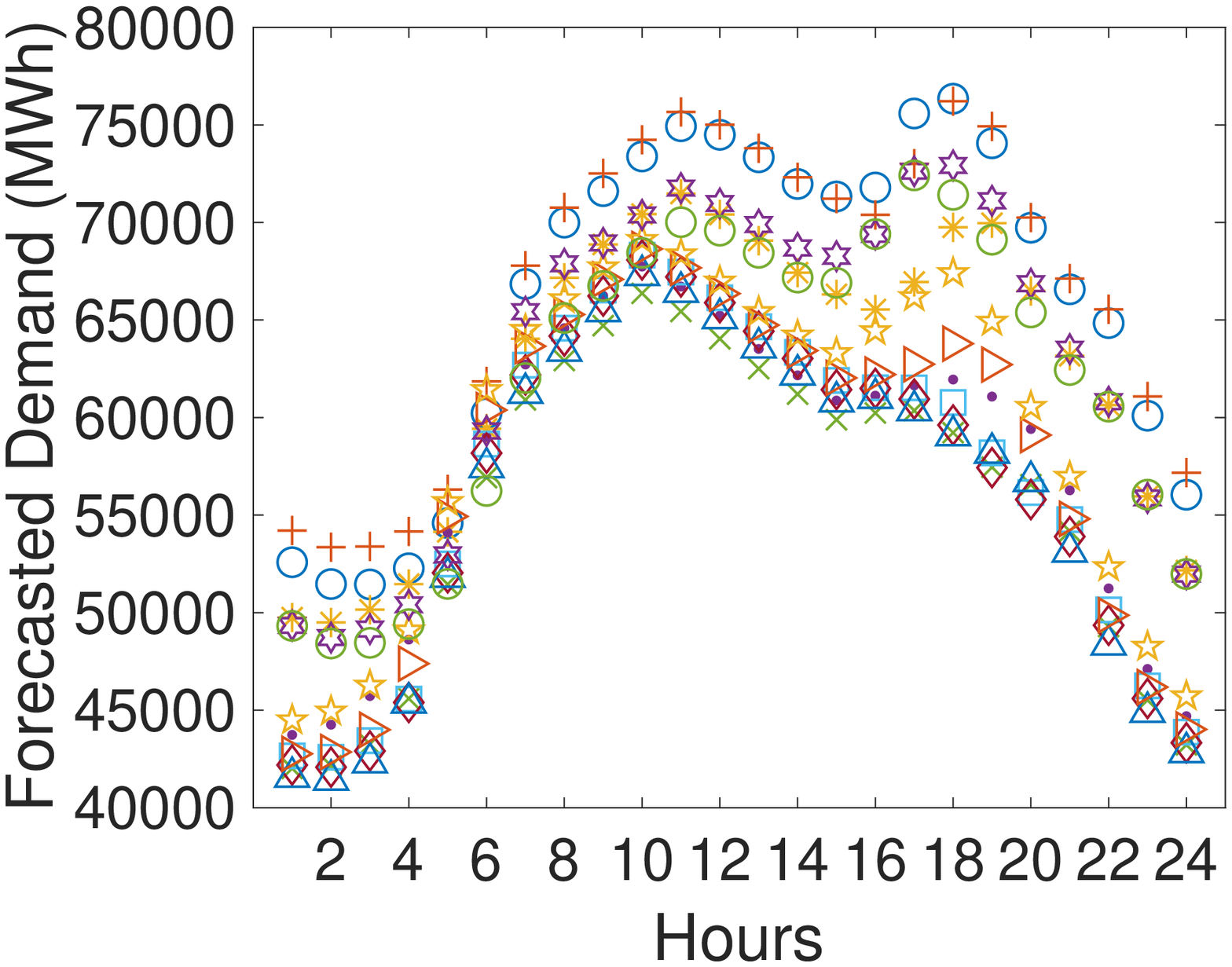}\\
\includegraphics[width=5cm]{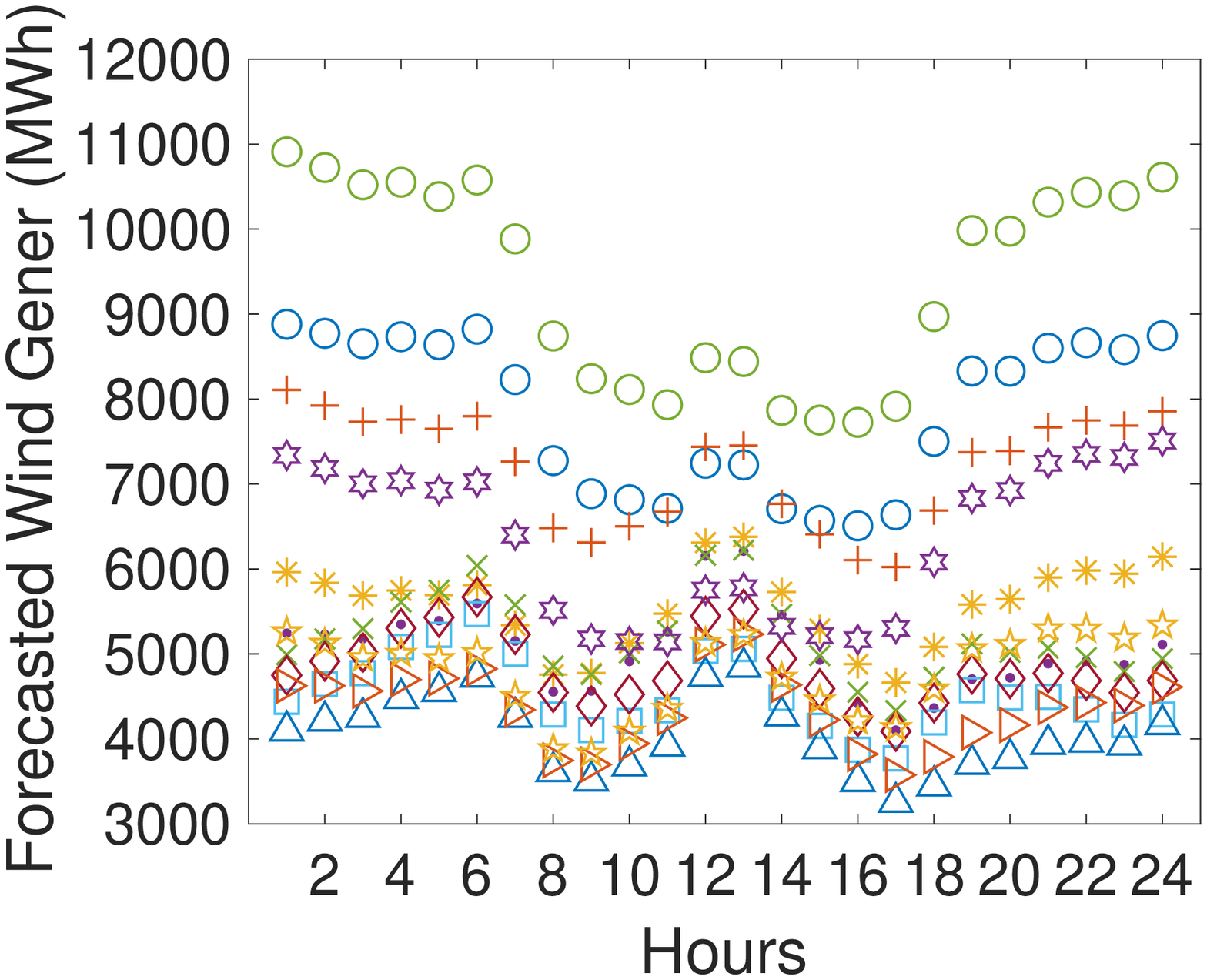}&
\includegraphics[width=5cm]{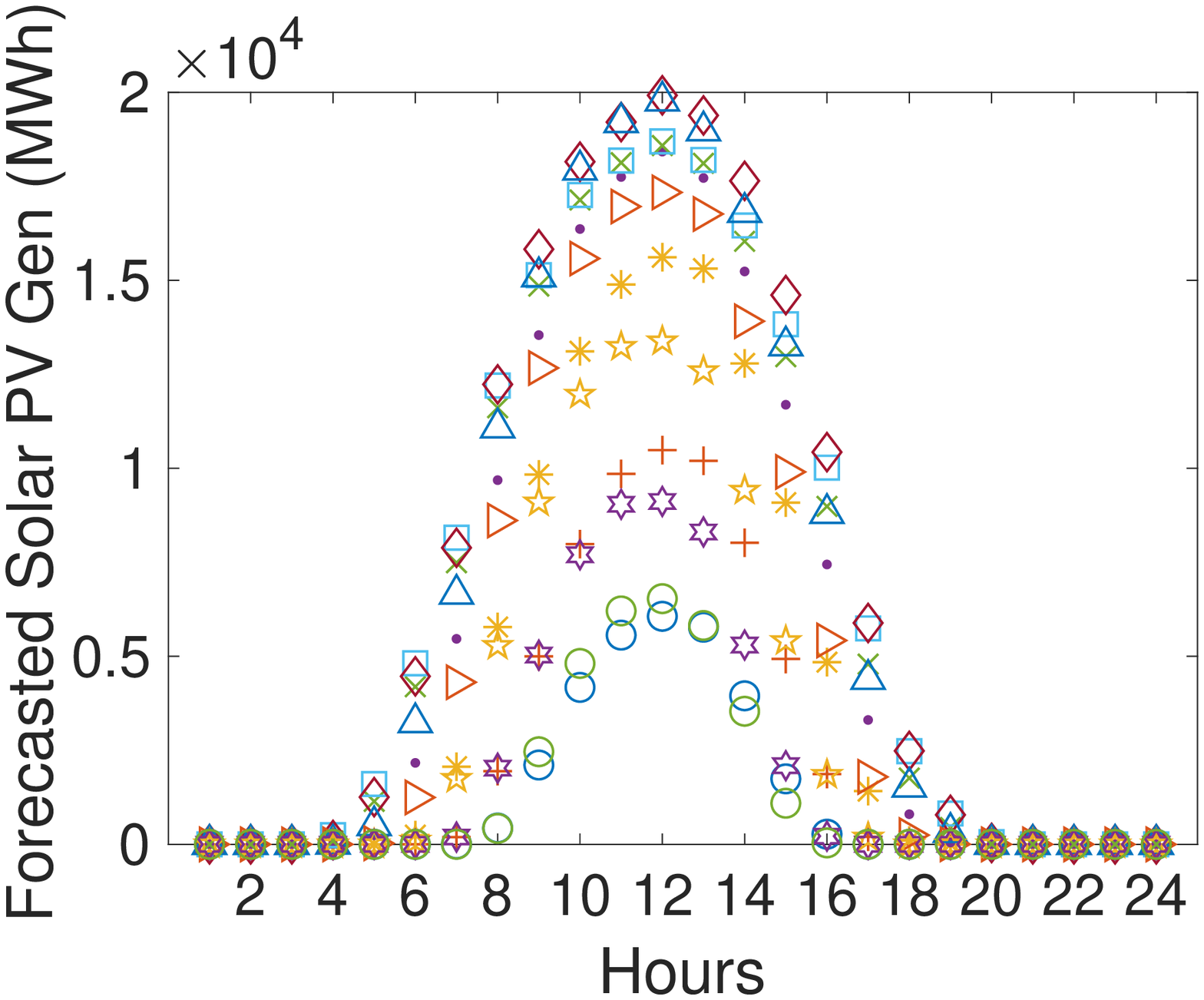}\\
\end{tabular}
\caption{\small{Intra-daily profiles of Monthly Averages for Electricity Day-ahead Prices (top left in \euro/MWh), Forecasted Demand (top right in MW), Forecasted Wind Generation (bottom left in MW), and Forecasted Solar PV Generation (bottom right in MW) observed in Germany. [January (blue $\circ$), February ($+$), March ($\star$), April ($\bullet$), May ($\times$), June ($\square$), July ($\diamond$), August ($\triangle$), September ($\triangleright$), October ($\pentagon$), November ($\hexagon$), December (green $\circ$)]}}% from 01/01/2011 to 31/12/2016.}}
\label{intra-daily-country-profiles-monthly}
\end{figure}

\begin{figure}[h!]
\centering
\begin{tabular}{cc}
% German intradaily profiles
\includegraphics[width=5cm]{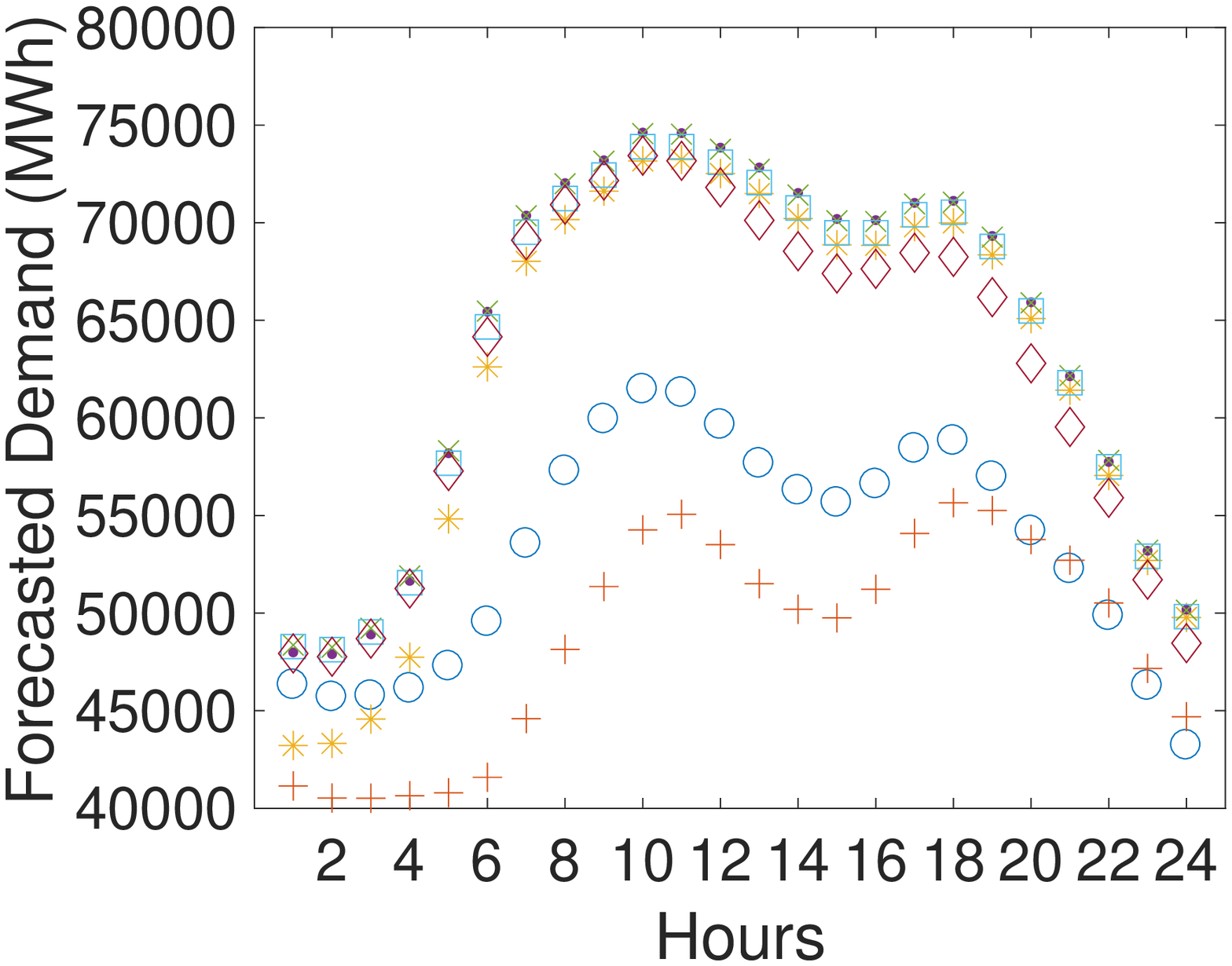}&
\includegraphics[width=5cm]{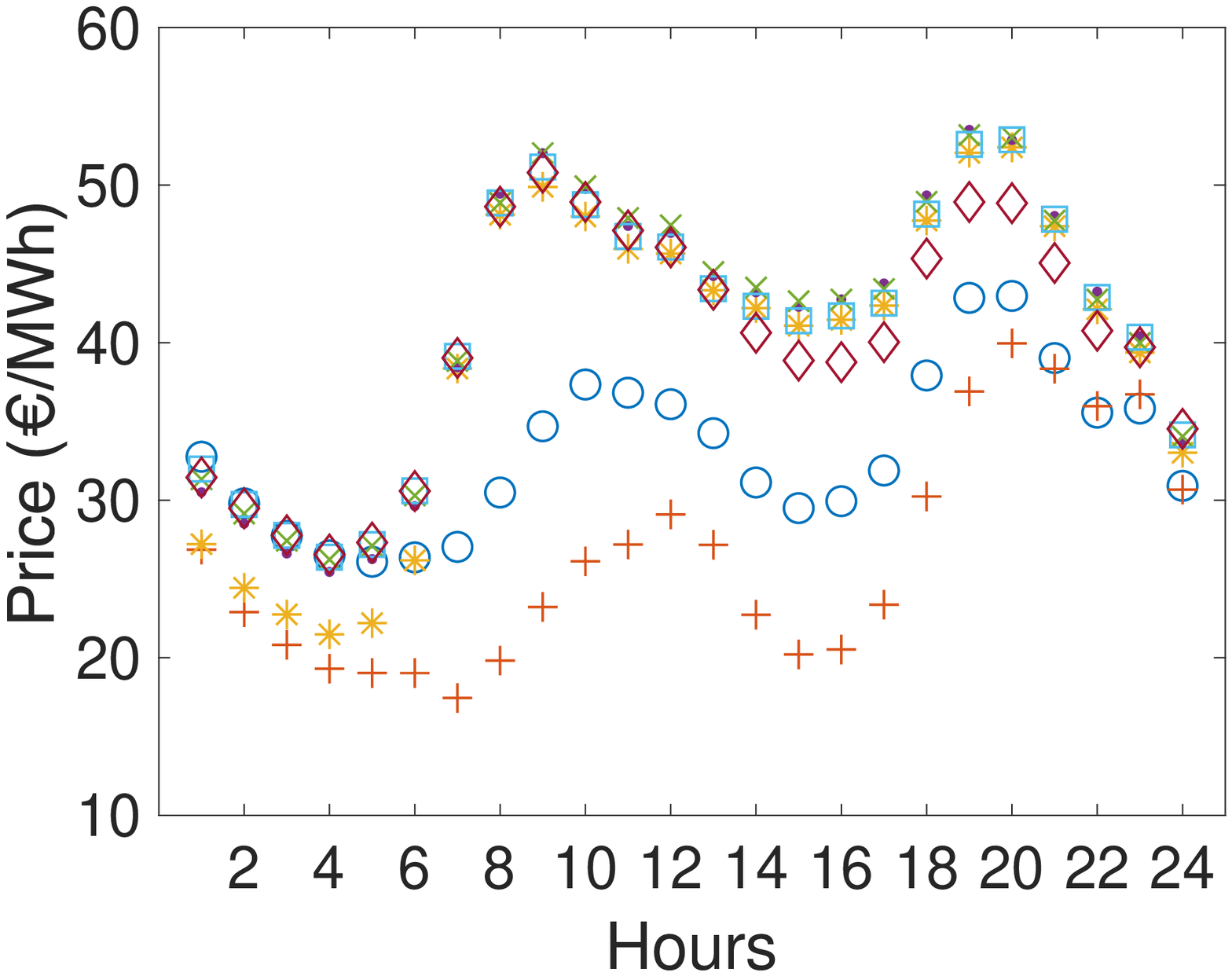} \\
\end{tabular}
\caption{\small{Intra-daily profiles across days of the week for Forecasted Demand (on the left) and Day-Ahead Electricity Prices (on the right) in Germany. [Saturday ($\circ$), Sunday ($+$), Monday ($\star$), Tuesday ($\bullet$), Wednesday ($\times$), Thursday ($\square$), Friday ($\diamond$)].}}
\label{intra-daily-country-profiles-dow}
\end{figure}

To summarize, we use daily fossil fuel prices ($\text{CO}_2$, gas, and coal, denoted by $m, g$, and $c$, respectively, and kept constant over the 24 hours) and hourly data (with daily frequency) for electricity prices, forecasted demand (denoted by $x$), wind (denoted by $w$), and solar PV generation (denoted by $z$) from 01 January 2011 to 31 December 2016 for Germany and Denmark and from 13 June 2014 to 13 June 2017 for Italy and Spain. We use the first four years as an estimation sample for Germany and Denmark, and the first two for Italy and Spain, whereas we use the last two/one years as the forecast evaluation period. The historical dynamics of these series observed in Germany are reported in Figure \ref{Plot_Germany}. % see label not working \ref{Supp_Graph}}
Prices show clearly the new stylized fact of ``downside'' spikes together with mean-reversion, whereas forecasted demand and solar generation exhibit more clear yearly seasonal patterns, with an increasing trend for solar power generation according to the new capacity additions through years.
Similarly, forecasted wind shows its dependence on weather conditions albeit with an increasing trend,\footnote{Given that trends can be observed in the studied series, we have tested that its inclusion does not improve substantially the forecasting performance.} corresponding again to investments in new capacity. To highlight calendar seasonality, monthly profiles for electricity prices, forecasted demand, and wind and solar generation are depicted in Figure \ref{intra-daily-country-profiles-monthly}. Furthermore, to emphasize the weekly seasonality, Figure \ref{intra-daily-country-profiles-dow} depicts the intra-daily dynamics across days of the weeks for demand and prices; obviously, wind and solar are not presented, as they are weather-dependent. Similar figures for the other countries are reported in Section S.1 of the Supplementary Material.


\begin{figure}[h!]
\centering
\begin{tabular}{ccc}
\includegraphics[width=5cm]{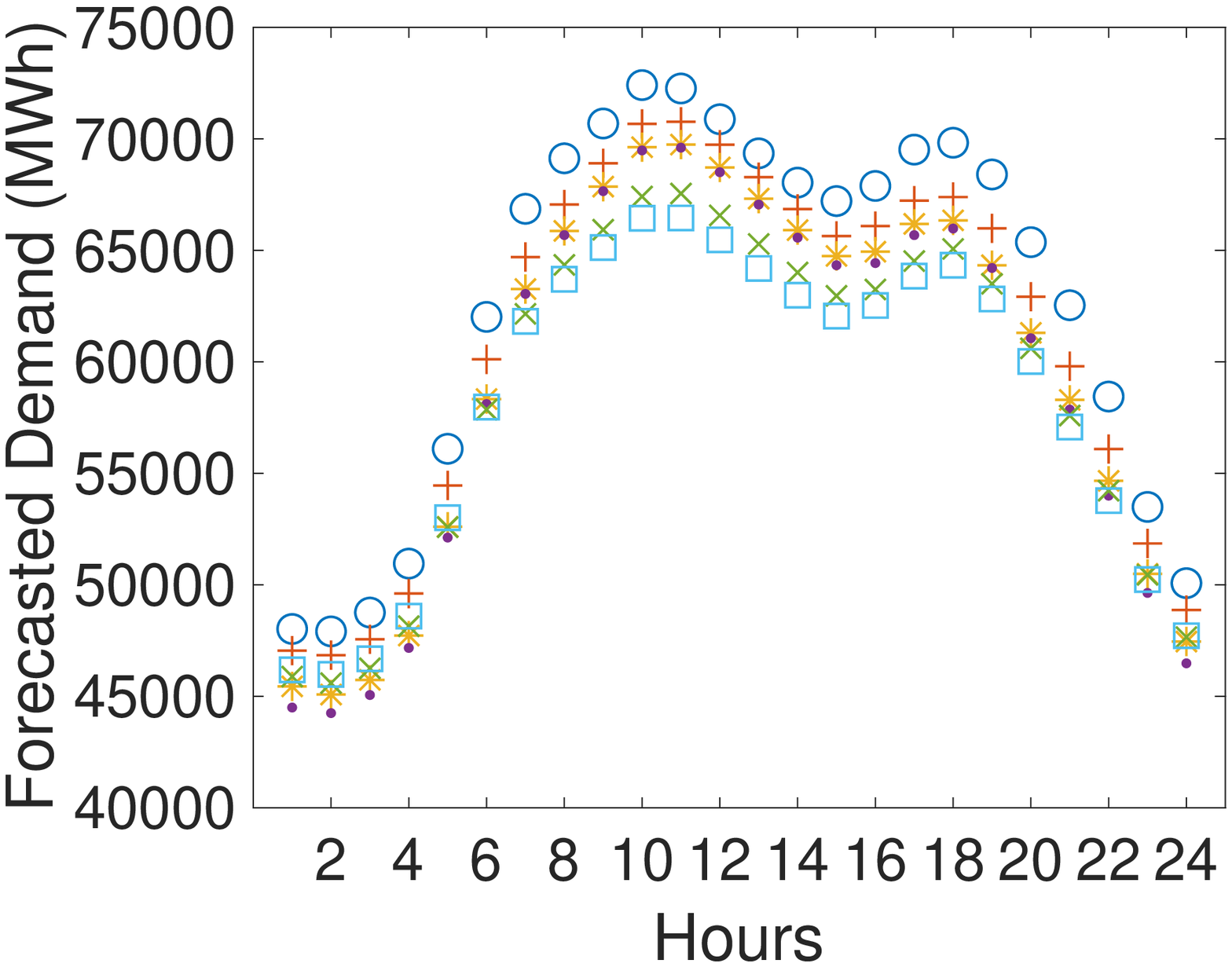} &
\includegraphics[width=5cm]{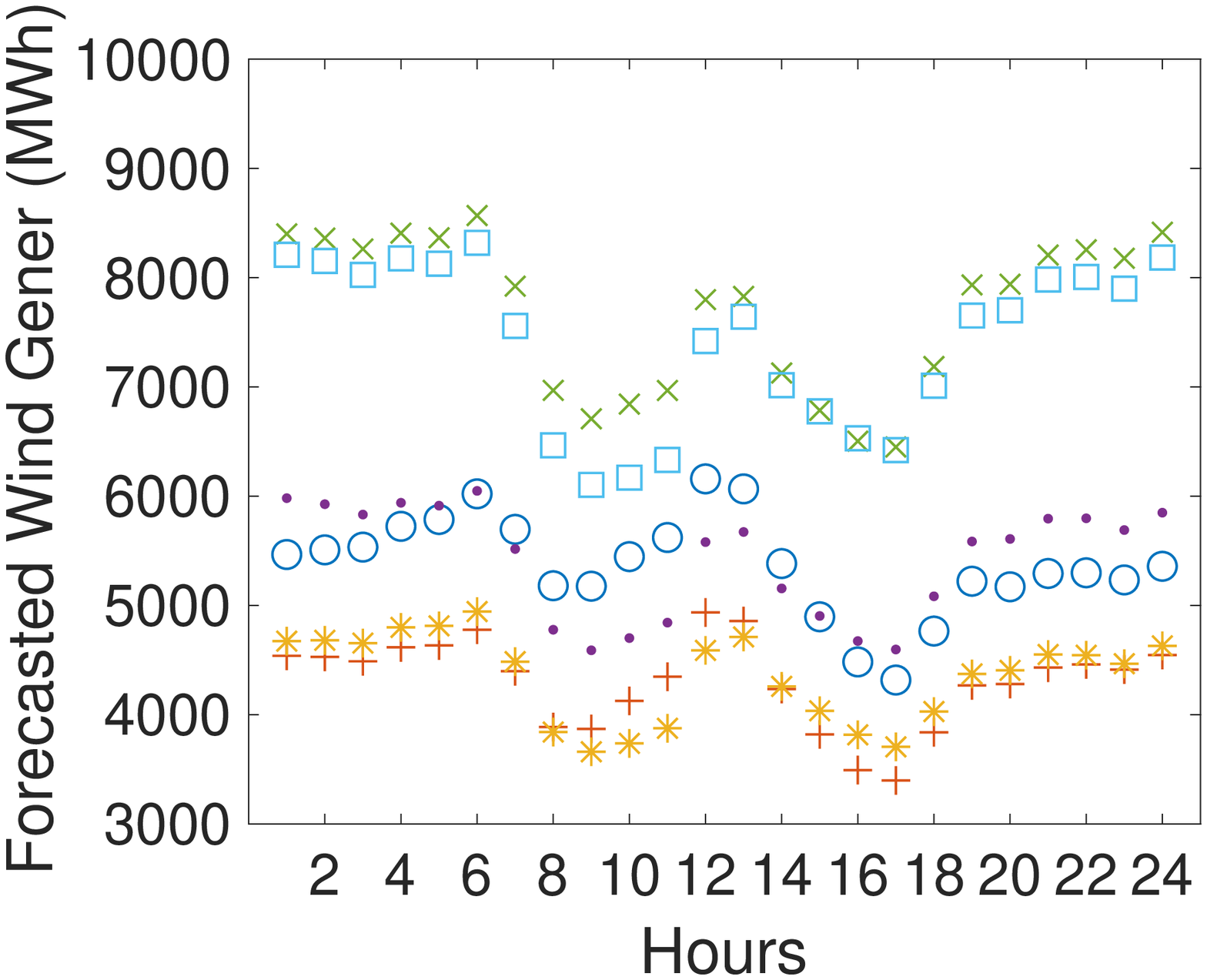}&
\includegraphics[width=5cm]{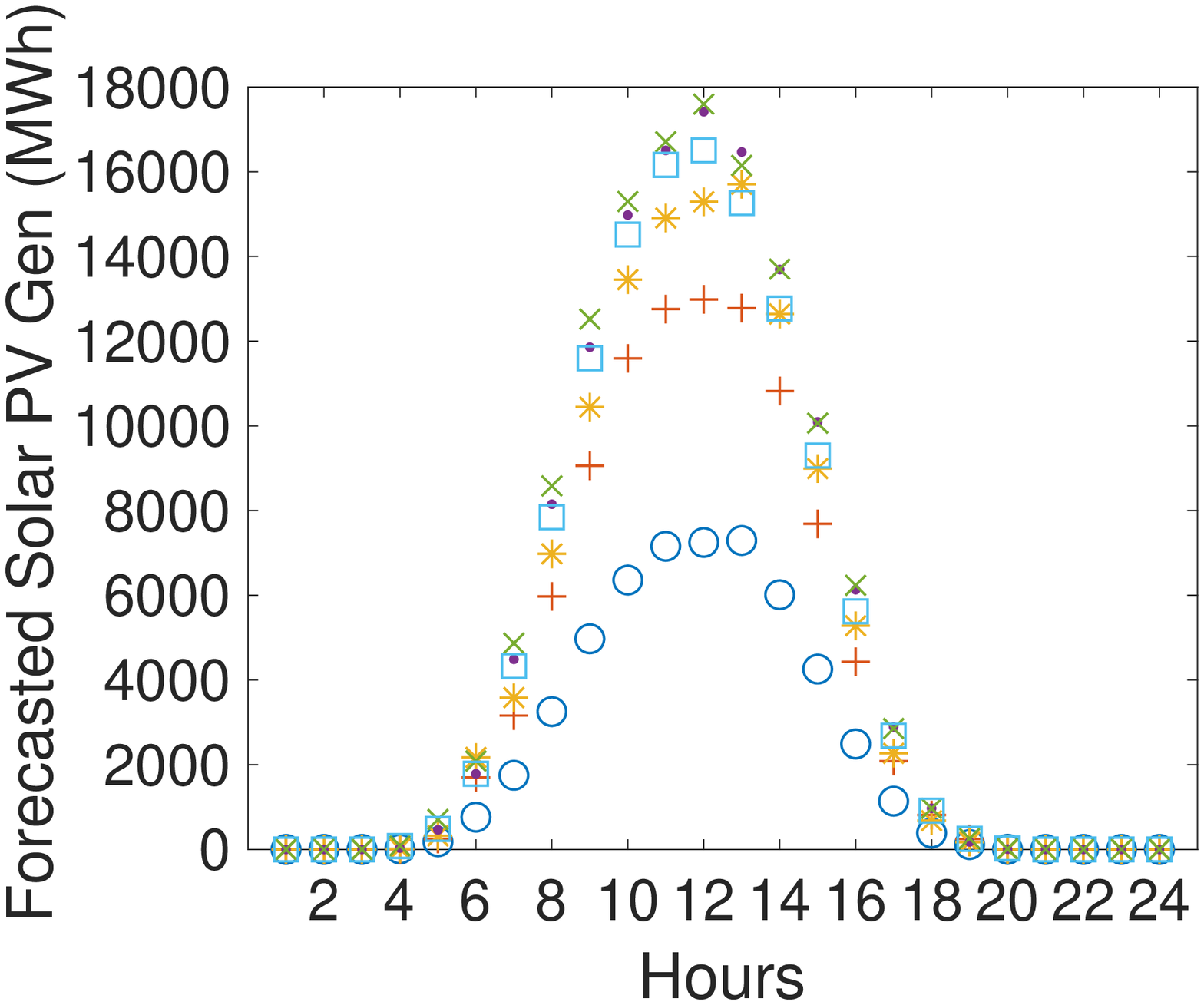}\\
\includegraphics[width=5cm]{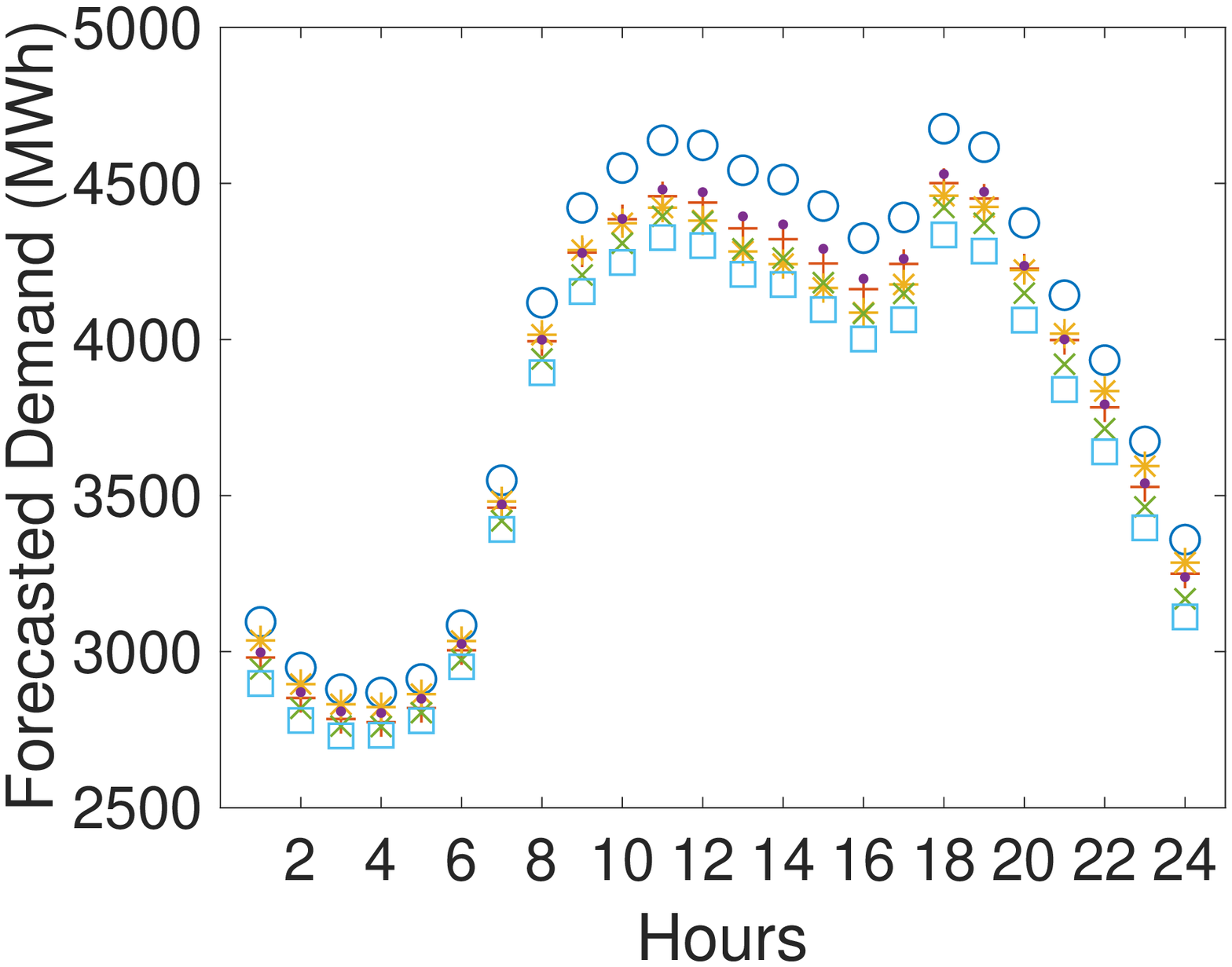} &
\includegraphics[width=5cm]{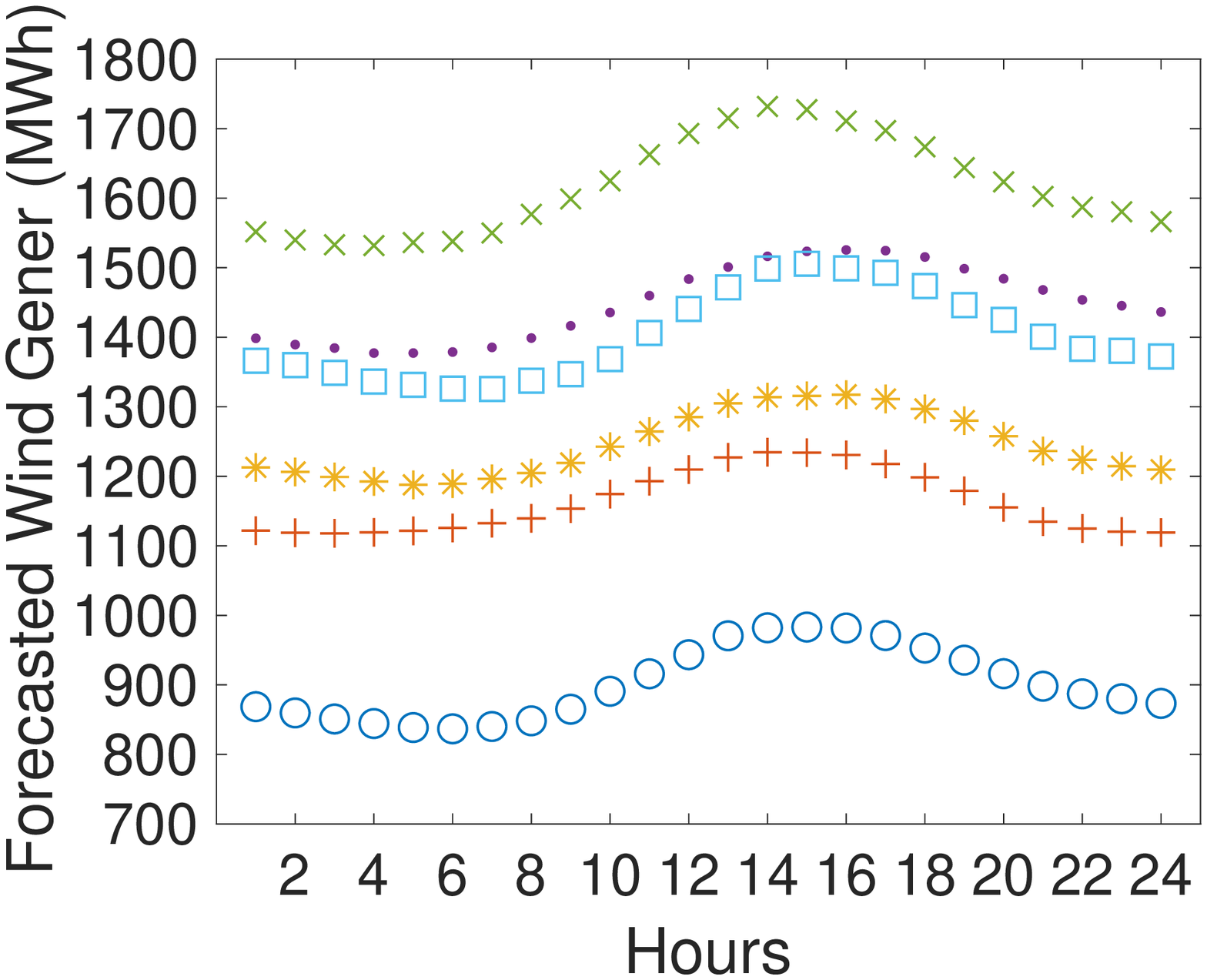}\\
\includegraphics[width=5cm]{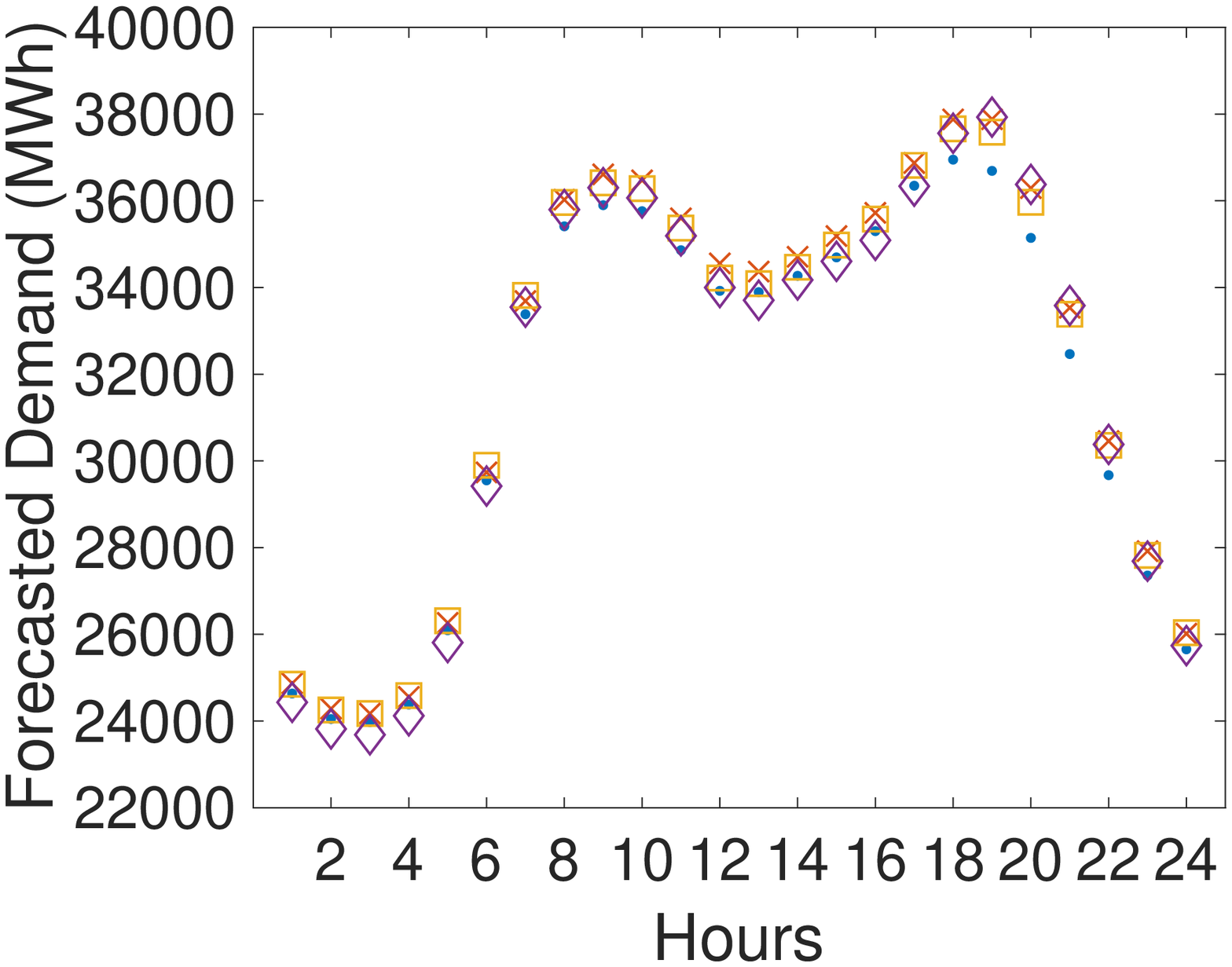} &
\includegraphics[width=5cm]{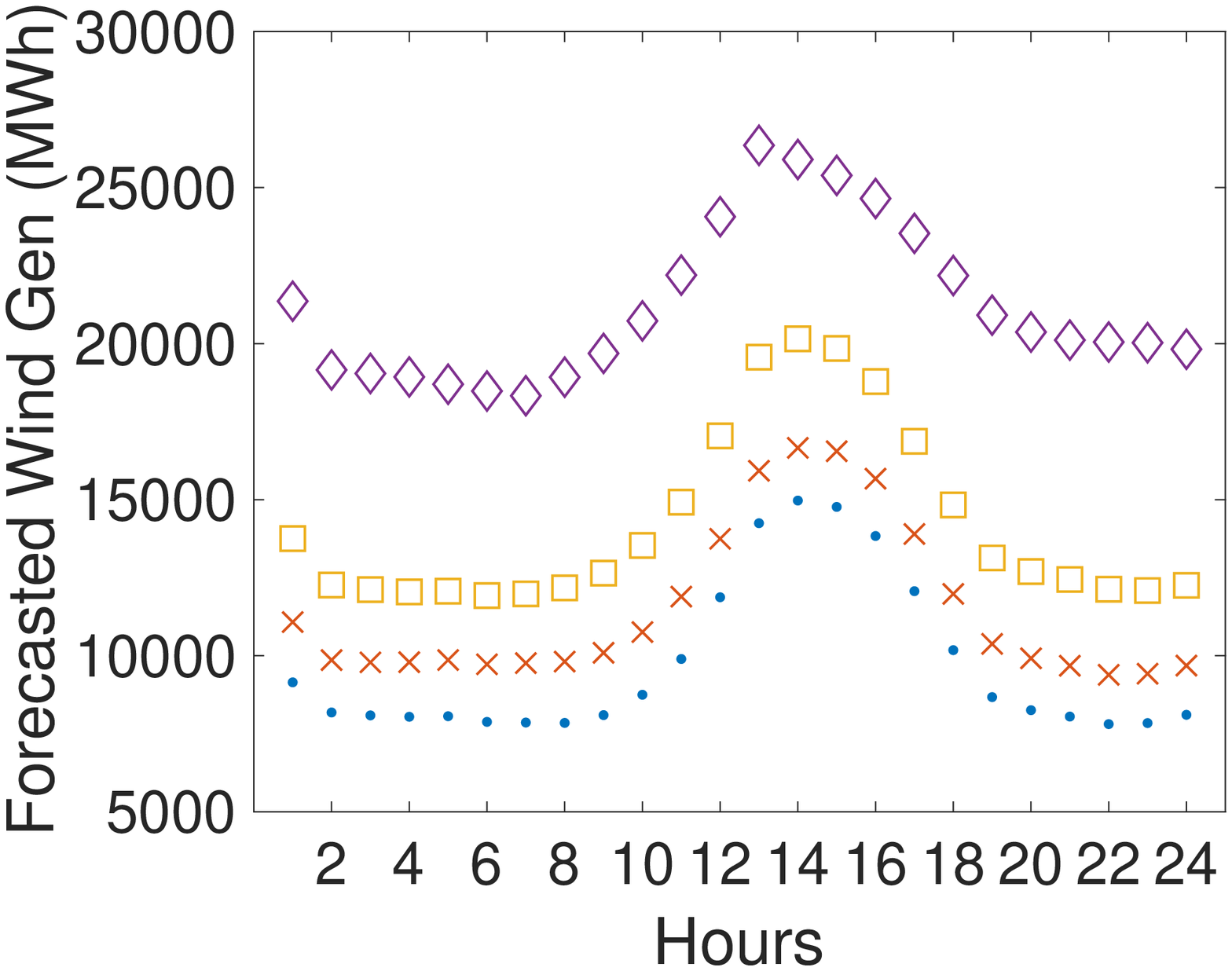}&
\includegraphics[width=5cm]{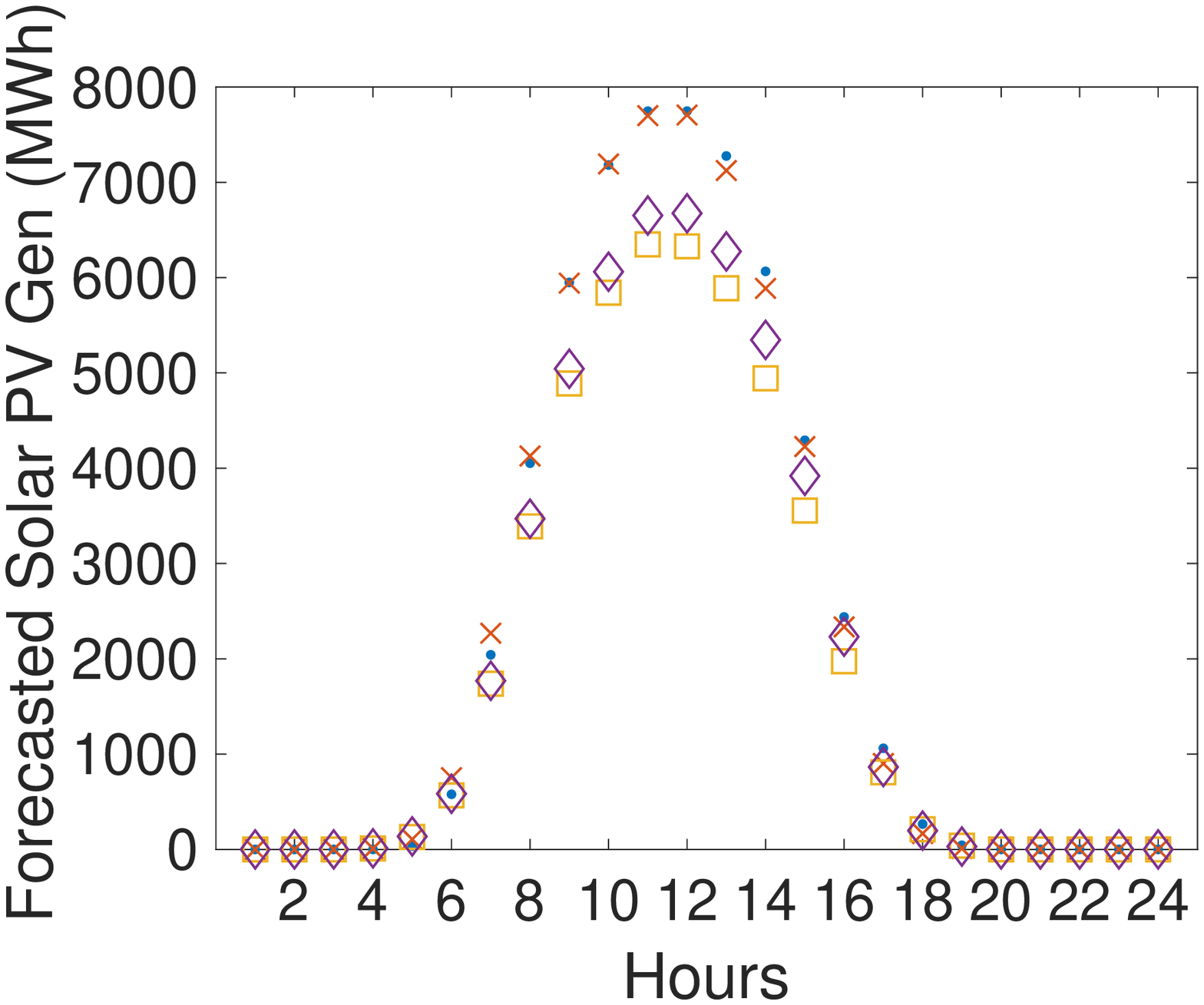}\\
\includegraphics[width=5cm]{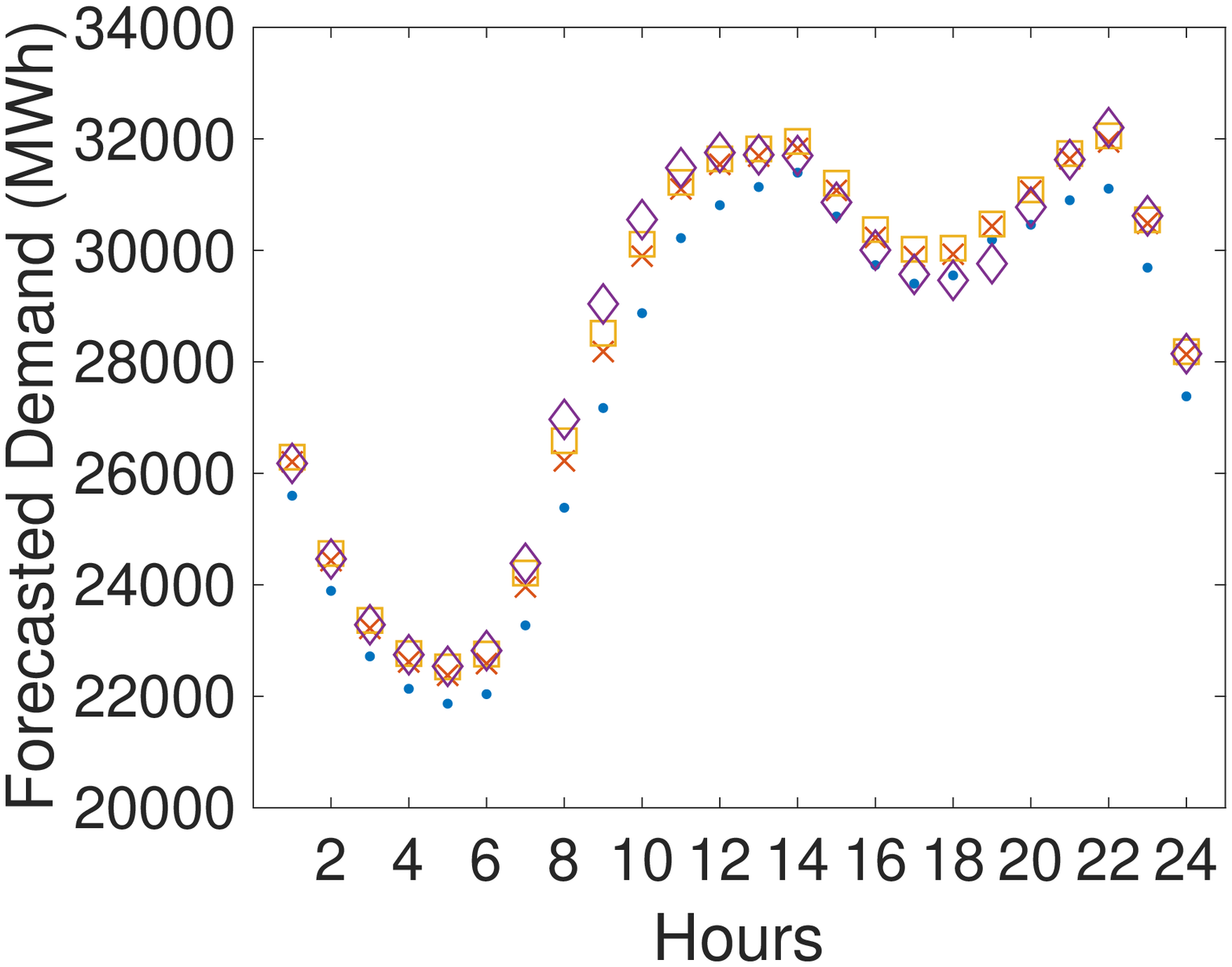} &
\includegraphics[width=5cm]{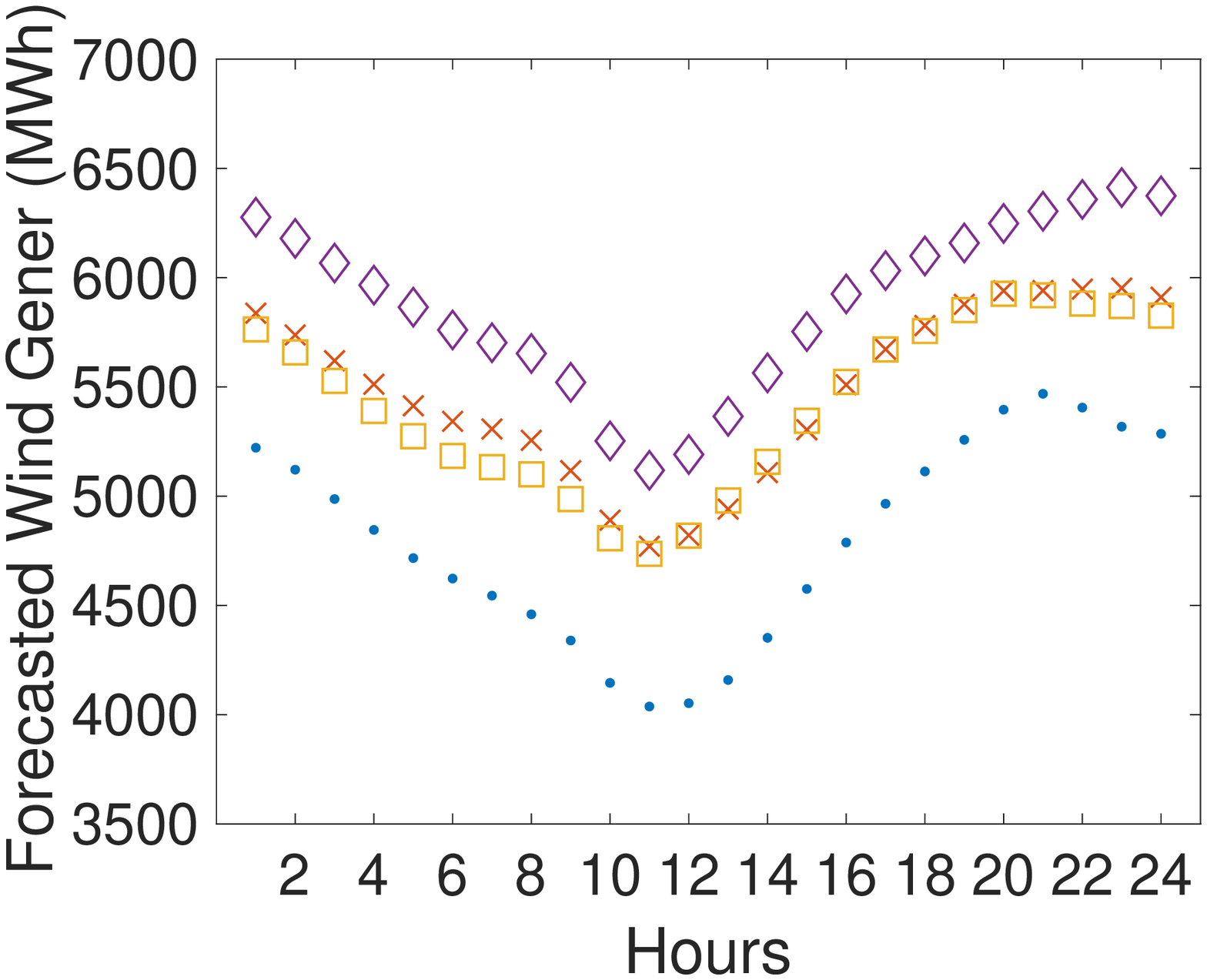}&
\includegraphics[width=5cm]{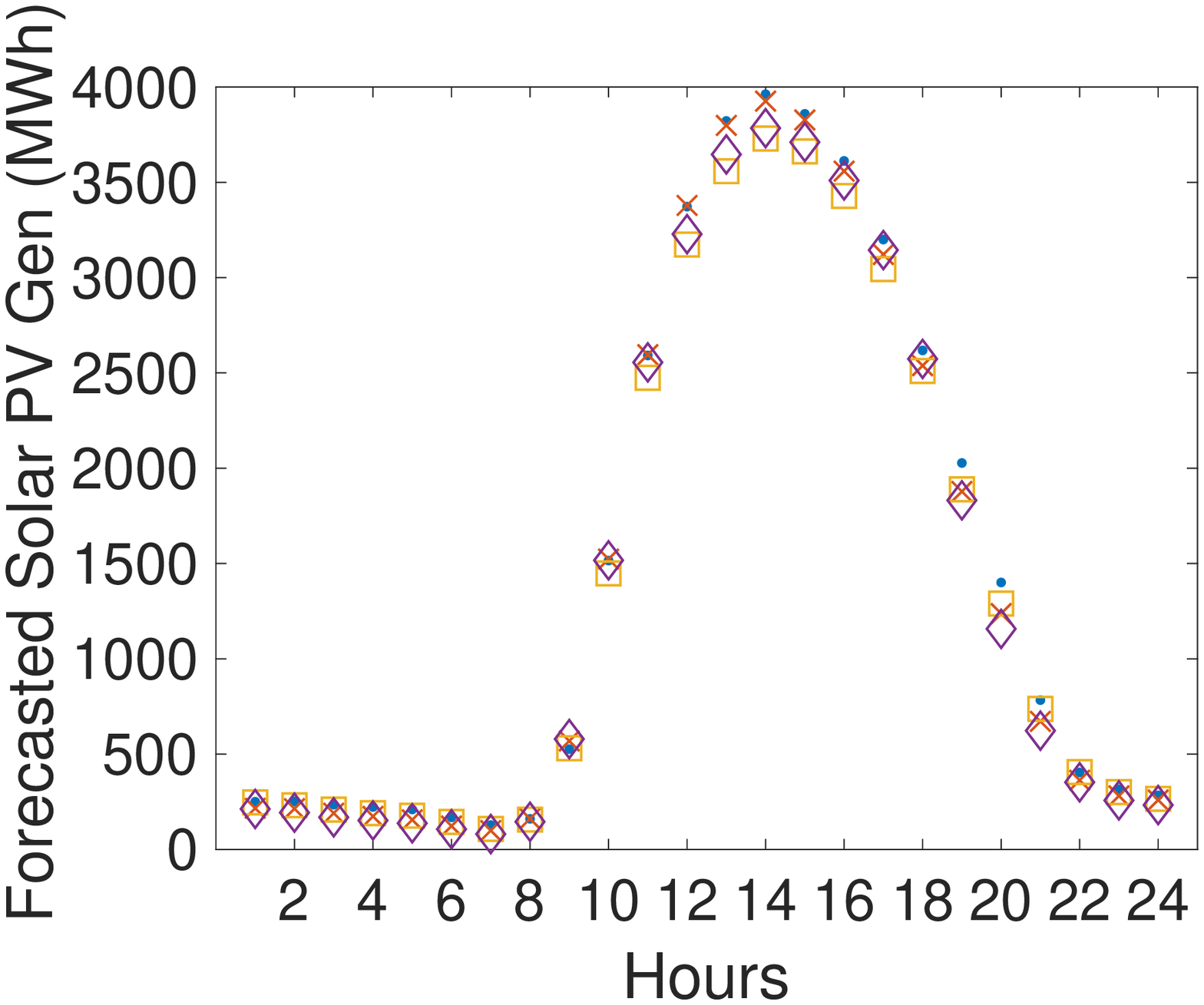}\\
\end{tabular}
\caption{\small{Intra-daily profiles of Yearly Averages for Forecasted Demand (left in MW), Forecasted Wind Generation (centre in MW), and Forecasted Solar PV Generation (right in MW) observed in Germany (first row), Denmark (second row), Italy (third row), and Spain (last row). [2011 (blue $\circ$), 2012 ($+$), 2013 ($\star$), 2014 ($\bullet$), 2015 ($\times$), 2016 ($\square$)]}}
\label{intra-daily-country-profiles-yearly}
\end{figure}

Finally, the intra-daily profiles for the yearly average values of forecasted demand and RES-E are represented in Figure \ref{intra-daily-country-profiles-yearly} to identify scenarios of high/low demand and/or RES-E expected to affect prices and, consequently, forecasts. %These results are presented in the Supplementary Material OR have been omitted for lack of space but are available on request.
We can observe that the ramp-up hours (during which the demand for electricity is expected to grow substantially) as well as the ramp-down hours (when demand is expected to decrease sharply) change across markets according to day- and night-time and geographical locations.
However, they confirm higher demand levels in the peak period (roughly between 8 a.m. and 8 p.m. for all markets). The intra-daily profiles for wind show different dynamics: we can again identify scenarios for high wind generation during peak hours in Denmark and Italy, whereas the opposite occurs in Germany and Spain.
Obviously, the intra-daily profiles for solar PV generation are, instead, common for all markets, where available. Therefore, we can expect a stronger combined effect of high demand and wind in Denmark, and high demand, wind, and solar in Italy, but contrasting scenarios for demand, wind, and solar during the day in Germany and Spain: a low-high-low one (that is low demand and solar versus high wind) in the early and late hours versus a high-low-high one (that is high demand and solar versus low wind) for peak hours.

%%%%%%%%%%%%%%%%%%%%%%%%%%%%%%%%%%%%%%%%%%%%%%%%%%%%%%
%%         		  		Models	   				  %%%
%%%%%%%%%%%%%%%%%%%%%%%%%%%%%%%%%%%%%%%%%%%%%%%%%%%%%%
\section{Forecasting Models}
\label{sec_Models}

We consider univariate and multivariate models for hourly prices with seasonality and with the introduction of exogenous variables relative to the forecasted demand and forecasted RES-E. Furthermore, we have included fossil fuels to account for marginal costs, hence reflecting the non-linearity of the supply curve.
Specifically, coal, natural gas, and CO$_2$ settlement prices have been included with a delay of one day, given that market operators do know their values determined at the market closure of the day before (that is on day $t-1$) when they run their models early in the morning on day $t$ to submit the 24-hour price forecasts by 11 a.m. (of the same day) for trades occurring on the following day, $t+1$.

Therefore, we have specified the following models to compare the forecasting performances when demand, RES-E, and fossil fuels are taken into account.
There is unanimous consensus that including demand forecasts or RES-E forecasts (if the market penetration is not negligible) leads to more accurate forecasts. However, it is still an open question as to which RES forecast is more informative in which market and also whether their inclusion can reduce the importance of fossil fuels; hence, models with only a subset of exogenous variables have also been considered.
All together results, first, in inspecting simple models with only dummy variables for seasonality; second, in adding regressors accounting for both demand and supply curves (that is dummies plus forecasted demand and lagged fossil fuels); third, in considering if the forecasting ability of demand and RES reduces the need of including fuels; and finally, in verifying if only forecasted wind and/or solar generation is/are efficient in providing good price forecasts.
We follow common practice in the literature and restrict lags to $t-1,\,t-2$ and $t-7$, which correspond to the previous day, two days before, and one week before delivery time, recalling, first, similar conditions that may have characterized the market over the same hours and similar days (like congestions and blackouts) and, second, the demand level during the days of the week. \cite{Knittel2005}, \cite{WeronMisiorek2008} and \cite{Ravivetal2015} show that these specifications provide accurate forecasts because they capture seasonal patterns in electricity prices. In addition, this formulation reduces the risk of overparameterization. Hence, hourly prices with a reduced 7-lag structure are considered, and, with an abuse of notation in the remainder of the paper, $p=3$ is used in all our univariate and multivariate models to denote the number of included lags, instead of the maximum lag.
\subsection{Multivariate Models}

We consider and compare the performances of two different multivariate model specifications with and without exogenous variables, used as benchmarks for the corresponding multivariate models. These are the VAR model, the VAR model with exogenous variables (VARX) estimating by using Least Square (OLS), see equations (2.3.2) and (2.3.4) in \cite{KilianLuktepol2017}, and their Bayesian formulations (BVAR and BVARX, respectively) with a normal-Wishart prior, see Section S.2 of the Supplementary Material.

\subsubsection{Vector Autoregressive Model -- VAR}

Let $\mathbf{y}_t = (y_{1t},\dots,y_{Ht})'$ denote the $(H\times 1)$ vector of hourly electricity prices, with $H = 24$. Moreover, we denote with $\mathbf{d}_t = (d_{1t}, \dots, d_{Kt})'$ the $(K\times 1)$ dummy vector with $(d_{1t},\dots,d_{12t})$ representing the twelve months of the year and $(d_{13t},d_{14t})$ representing Saturdays and Sundays, hence $K = 14$. The VAR model of order $p$ is formulated as follows:
\begin{equation}
\mathbf{y}_t = \Phi' X_t + \mathbf{e}_t, \quad t=1,\dots,T, \label{VAR}
\end{equation}
where $\Phi$ is the $((Hp +K) \times H)$ matrix containing the autoregressive coefficients as well as the coefficients for all dummy variables, and $X_t = (\mathbf{y}_{t-1},\dots, \mathbf{y}_{t-p}, \mathbf{d}_t)$ is the matrix $((Hp +K) \times H)$ made by the lagged electricity prices and the dummy variables. The vector of errors $\mathbf{e}_t$ is assumed to be serially uncorrelated and normally distributed with zero mean and a full covariance matrix $\Sigma$.

\subsubsection{Vector Autoregressive Model with Exogenous Variables -- VARX}

The VARX includes the forecasted demand, as well as the forecasted wind and solar power generation, when available, and fossil fuel prices for coal, gas, and $\text{CO}_2$. The exogenous demand and RES variables are represented by the following vectors of dimensions $(H\times 1)$, $\mathbf{x}_t = (x_{1t},\dots,x_{Ht})'$, $\mathbf{z}_t = (z_{1t},\dots,z_{Ht})'$ and $\mathbf{w}_t = (w_{1t},\dots,w_{Ht})'$, respectively. On the other hand, fuel prices do not change over the 24 hours and are determined on the previous day, $t-1$. Thus, $m_{t-1}$, $g_{t-1}$ and $c_{t-1}$ are the representations for $\text{CO}_2$, gas, and coal at previous time, respectively. From \eqref{VAR}, we re-define the matrix $X_t$ as
{
%\begin{equation*}
$X_t = \left(\mathbf{y}_{t-1},\dots, \mathbf{y}_{t-p}, \mathbf{d}_t, \mathbf{x}_{t}, \mathbf{z}_{t}, \mathbf{w}_{t}, m_{t-1}, g_{t-1}, c_{t-1}\right)$
%\end{equation*}
}
and, consequently, the matrix of coefficients $\Phi$ of size $((Hp +K + 3H +3) \times H)$. The matrix $X_t$ now comprises the vector of lagged hourly electricity prices, the vectors of dummy variables, and the exogenous variables.
From eq. \eqref{VAR}, as the observations vary with time $t=1,\dots,T$, the VAR and VARX models of order $p$ can be rewritten in a compact way
\begin{equation}
\mathbf{Y} = \mathbf{X} \Phi + \mathbf{E}, \label{VARX1}
\end{equation}
where $\mathbf{Y} = (\mathbf{y}_1',\dots, \mathbf{y}_T')$ is an $(T \times H)$ matrix, and $\mathbf{X} = (X_1,\dots,X_T)'$ is the $(T \times (Hp+K+3H+3))$ matrix of explanatory variables containing all the exogenous variables.\footnote{We have also performed the forecasting exercises including the lags (1,2, and 7) for exogenous variables, but the results were unchanged although computationally intensive and time-demanding. For these reasons, and having proper forecasts, we prefer to adopt the former models without lagged exogenous variables.} The $(T \times H)$ error matrix $\mathbf{E} = (\mathbf{e}_1',\dots, \mathbf{e}_T')$ is normally distributed and serially uncorrelated with covariance matrix $\Sigma$.
\subsubsection{Bayesian Vector Autoregressive Models -- BVARs}
Our multivariate models with or without exogenous variables have been additionally estimated using the Bayesian methodology. From eq. \eqref{VARX1}, a BVAR or BVARX has the following stacked form
\begin{equation}
\mathbf{y} = (I_{H} \otimes \mathbf{X}) \bm{\alpha} + \bm{\varepsilon}, \label{BVAR}
\end{equation}
where $\bm{\alpha} = \mbox{vec}(\Phi)$, $\mathbf{y} = \mbox{vec}(\mathbf{Y})$ are vectorized matrices, $\bm{\varepsilon} \sim \mathcal{N}(\mathbf{0}, \Sigma \otimes I_T),$ with $I_T$ being a $T$-dimensional identity matrix. This stacked form representation allows us to define and study the prior and posterior distribution of the matrix of coefficients and covariance matrix leading to a closed form distribution.
In particular, we define prior information on the matrix of coefficients and on the covariance matrix using a conjugate normal-Wishart prior.\footnote{We have performed the analysis using both a standard Minnesota and the normal-Wishart priors, and the results are similar. Therefore, due to lack of space, we have reported only the results for the latter. Details on the prior information and posterior distribution are reported in Section S.2 of the Supplementary Material.}
\subsection{Univariate Models}
For all previous models, we formulate 24 (parsimonious) univariate AR specifications with the same assumptions on the lag order of the VAR specifications, whereas the errors are assumed to be normally distributed with zero mean and $\sigma_h^2$ variance for the hours $h=1,\cdots,24$. The autoregressive model with only dummy variables is used as benchmark in the forecasting comparisons and can be written as follows
\begin{equation}
y_{h,t} = \sum_{l=1}^p \phi_l y_{h,t-l} + \sum_{k=1}^K \psi_k d_{kt} +\varepsilon_{h,t}. \notag
\end{equation}
On the other hand, the univariate ARX or BARX can be written as
\begin{equation*}
y_{h,t} = \sum_{l=1}^p \phi_l y_{h,t-l} + \sum_{k=1}^K \psi_k d_{kt} + \alpha_1 x_{ht} + \alpha_2 z_{ht} + \alpha_3 w_{ht} + \beta_1 \text{m}_{t-1} + \beta_2 \text{g}_{t-1} + \beta_3 \text{c}_{t-1} + \varepsilon_{h,t}
\end{equation*}
%\begin{align*}
%y_{h,t} &= \sum_{l=1}^p \phi_l y_{h,t-l} + \sum_{k=1}^K \psi_k d_{kt} + \alpha_1 x_{ht} + \alpha_2 z_{ht} + \alpha_3 w_{ht} \\
%&+ \beta_1 \text{c}_{t-1} + \beta_2 \text{g}_{t-1} + \beta_3 \text{l}_{t-1} + \varepsilon_{h,t},
%\end{align*}
where $x_{ht}, z_{ht}$ and $w_{ht}$ represent (forecasted) demand and renewable energy variables, whereas $m_{t-1}, g_{t-1}$ and $c_{t-1}$ are the fossil fuel prices previously described. Even in the univariate case, we use both the frequentist and the Bayesian estimation procedures.

To support the multivariate formulation, we run $24$ univariate models with dummies, lags of $y_t$, and fundamentals lagged same-hour prices, adding also the first lag of all other remaining hours, that is
\begin{equation*}
y_{h,t} = \sum_{l=1}^p \phi_l y_{h,t-l} + \sum_{k=1}^K \psi_k d_{kt} + \alpha_1 x_{ht} + \alpha_2 z_{ht} + \alpha_3 w_{ht} + \beta_1 \text{m}_{t-1} + \beta_2 \text{g}_{t-1} + \beta_3 \text{c}_{t-1} +  \sum_{j\ne h} \gamma_j y_{j,t-1} + \varepsilon_{h,t}
\end{equation*}
Then, from the resulting 24 residual series of each model, $\hat{\varepsilon}_{h,t}$, the variance-covariance matrix has been computed. Uncorrelated residuals make the multivariate VAR specification unnecessary; however, we find evidence of large correlations across all studied markets\footnote{These results have been omitted for lack of space, but they are available on request.}. Therefore, a VAR with full covariance matrix $\Sigma$ seems more appropriate to estimate this covariance structure and it should result in improved density forecast accuracy.

\subsection{Forecast Assessment}
We assess the goodness of our forecasts using different point and density metrics. Considering the accuracy of point forecasts, we use the root mean square errors (RMSEs) for each of the hourly prices, as well as the RMSEs on the daily average and on an average restricted only to central hours, as specified below. The RMSE for $h=1,\dots, 24$ hourly prices is computed as
\begin{equation}
\mbox{RMSE}_h = \sqrt{\frac{1}{T-R} \sum_{t=R}^{T-1} \left(\hat{y}_{h,t+1|t} - y_{h,t+1} \right)^2}, \label{RMSE}
\end{equation}
where  $T$ is the number of observations, $R$ is the length of the rolling window and $\hat{y}_{h,t+1|t}$ are the individual hourly price forecasts. In addition, we analyse the average RMSEs on all the 24 hours ($\mbox{RMSE}_{\text{Avg}}$) and on the hours from 8 a.m. to 8 p.m. (peak hours, $\mbox{RMSE}_{\text{Avg}}^{P}$), computed as follows:
\begin{align}
\text{RMSE}_{\text{Avg}} &= \frac{1}{24}\sum_{h=1}^{24} \mbox{RMSE}_h, \label{RMSE_Avg} \\
\mbox{RMSE}_{\text{Avg}}^{P} &= \frac{1}{13} \sum_{h=8}^{20} \mbox{RMSE}_h. \label{RMSE_Avg_Red}
\end{align}
To evaluate density forecasts, we use the average continuous ranked probability score (CRPS).\footnote{We computed also that the log predictive score and results were similar; hence, they have not been reported.}

As indicated in \cite{Gneiting2007} and \cite{GneitingRanjan2011}, some researchers view the continuous ranked probability score as having advantages over the log score. In particular, the CRPS does a better job of rewarding values from the predictive density that are close to - but not equal to - the outcome, and it is less sensitive to outlier outcomes. The CRPS, defined such that a lower number is a better score, is given by
\begin{equation}
\mbox{CRPS}_{h,t}(y_{h,t+1}) = \int_{-\infty}^{\infty} \left(F(z) - \mathbb{I}\{y_{h,t+1} \le z\}\right)^2 dz = E_f|Y_{h,t+1} -y_{h,t+1}| - 0.5 E_f|Y_{h,t+1} -Y'_{h,t+1}|, \notag %\label{CRPS}
\end{equation}
where $F$ denotes the cumulative distribution function associated with the predictive density $f$, $\mathbb{I}\{y_{h,t+1} \le z\}$ denotes an indicator function taking the value $1$ if $y_{h,t+1} \le z$ and $0$ otherwise, and $Y_{h,t+1}$ and $Y'_{h,t+1}$ are independent random draws from the posterior predictive density. In the same way we can construct the average CRPS over the 24 hours and over peak hours on day $t+1$.

More specifically, we report the RMSEs and average CRPS for all the univariate and multivariate models and for every third hour.\footnote{Tables with all hours are available in Section S.4, S.5, S.6 and S.7 of the Supplementary Material.} %For the other AR (VAR) models, we report the ratios computed between the RMSE of the current models and the RMSE of the baseline AR (VAR) model. Then, entries of less than 1 indicate that the given current model yields forecasts more accurate than are those provided from the baseline, and similarly for the CRPS.

%More specifically, we report the RMSEs and average CRPS for both the baseline AR and VAR models and for every third hour.\footnote{Tables with all hours are available in Section \ref{Supp_Tab_Ger}, \ref{Supp_Tab_Den}, \ref{Supp_Tab_Ita}, and \ref{Supp_Tab_Spa} of the Supplementary Material.} For the other AR (VAR) models, we report the ratios computed between the RMSE of the current models and the RMSE of the baseline AR (VAR) model. Then, entries of less than 1 indicate that the given current model yields forecasts more accurate than are those provided from the baseline, and similarly for the CRPS.

In addition, %to provide a rough gauge of whether the differences in forecast accuracy are significant,
we apply \cite{Diebold1995} t-tests for equality of the average loss (with loss defined as squared error or CRPS) to compare predictions of alternative models to the benchmark for a given horizon $h$\footnote{In our application for testing density forecasts, we use equal weights without adopting a weighting scheme, as in \cite{AmisanoGiacomini2007}.}. The differences in accuracy that are statistically different from zero are denoted with one, two, or three asterisks, corresponding to significance levels of 10\%, 5\%, and 1\%, respectively. The underlying p-values are based on t-statistics computed with a serial correlation-robust variance, using the pre-whitened quadratic spectral estimator of \cite{Andrews92}. Our use of the Diebold-Mariano test, with forecasts from models that are, in many cases, nested, is a deliberate choice, as in \cite{ClarkRav15}, and, as noted by \cite{Clark2007} and \cite{Clark2012}, this test is conservative and might result in under-rejection of the null hypothesis of equal predictability. We report p-values based on one-sided tests, taking the AR (VAR) as the null and the other current models as the alternative.

Finally, we have also applied the Model Confidence Set procedure of \cite{hansen_etal.2011} across models for a fixed horizon to jointly compare their predictive power without disentangling between univariate and multivariate models. The \textsf{R} package \textsf{MCS} detailed in \cite{bernardi_catania.2016} has been used, and the differences have been tested separately for each hour and model, repeating the full process across all countries. Results are discussed in the following section.
%
%%%%%%%%%%%%%%%%%%%%%%%%%%%%%%%%%%%%%%%%%%%%%%%%%%%%%%
%%         		  	Results		   				  %%%
%%%%%%%%%%%%%%%%%%%%%%%%%%%%%%%%%%%%%%%%%%%%%%%%%%%%%%
\section{Results}
\label{sec_Results}
Our results are based on a one-step-ahead forecasting process with a rolling window approach of 4 years for Germany and Denmark and of 2 years for Italy and Spain. Let us recall that we have two estimation samples 01/01/2011--31/12/2014 for Germany and Denmark, and 13/06/2014--13/06/2016 for Italy and Spain. And then we have two forecast evaluation periods: 01/01/2015--31/12/2016 for the former two markets (for a total of 731 observations), and 14/06/2016--13/06/2017 for the latter two countries (hence, only 365 observations).
%Hence, we have two different forecast evaluation periods: the last two years from 01 January 2015 to 31 December 2016 (for Germany and Denmark, hence 731 observations), whereas only the last year from 14 June 2016 to 13 June 2017 for Italy and Spain (hence, only 365 observations given the unavailability of older historical data). %%Furthermore, for the multivariate models, we also run a model `restricted' to only the peak hours in which we have both forecasted wind and solar power generation.

Before evaluating the out-of-sample results, our in-sample evidence provides statistically significant coefficients for the RES variables in all markets; hence, confirming the empirical findings in previous literature on univariate models augmented with RES variables and extending similar conclusions also to multivariate models. In particular, coefficients of wind and solar are negative in Germany, Italy, and Spain. Also in Denmark, wind has a negative coefficient.\footnote{Detailed in-sample results are available under request.} These results confirm that renewable energy sources are significantly connected to and reduce electricity prices. Therefore, we continue our analysis by investigating whether these relationships can result in forecast gains.

To this end, our results show the performance of our different univariate and multivariate models from the simplest ones (with only dummy variables, the benchmarks) to more complex ones containing gas, coal, $\text{CO}_2$, and forecasts for demand, wind, and solar. Alternative formulations referred to subsets of drivers are described in Table S.1 of the Supplementary Material, and results are summarized in Tables \ref{table:RMSE_Rol} and \ref{table:CRPS_Rol}  across all markets at every third hours; whereas extensive comparisons are shown in Tables  S.2--S.9 of the Supplementary Material. %for both the RMSEs and CRPSs as well as for the DM tests and MCS. \textbf{-- va bene scritto cosi'?} % for Germany, Italy and Spain; whereas, no solar PV power is considered for Denmark.

Recalling the main objectives of this analysis, we have shown clearly the superiority of multivariate models when the full structure of 24 hours is considered. Multivariate VAR models outperform simple AR models when only seasonality is included. This holds true systematically across all countries, and according to both point and density metrics. For instance, in Germany, the average RMSE moves from 8.259 \euro/MWh in the univariate case to 6.839 \euro/MWh in the multivariate case. In Spain, it goes from 6.299 \euro/MWh to 5.110 \euro/MWh. The case is similar for the CRPSs for which we observe substantial reductions of almost 18\% (from 4.427 to 3.643) in Germany, 13\% (from 4.901 to 4.273) in Denmark, 5\% (from 3.658 to 3.469) in Italy, and 20\% (from 3.517 to 2.831) in Spain, for average values computed over the 24 hours. The AR models are included in the model confidence set in only 10 cases over 64 horizons in the two Tables \ref{table:RMSE_Rol} and \ref{table:CRPS_Rol} for both metrics and mainly for the Italian market. VAR models have a much higher frequency of inclusion. Hence, this supports our expectations of more efficient forecasts obtained considering the interrelationships among the whole 24 hours, as suggested by \cite{Stock2002} and anticipated by \cite{Ravivetal2015}.

Moreover, considering the most important fact, that is the forecasting improvements to the inclusion of RES and/or a subset of drivers, our results show that the Bayesian multivariate models with forecasted RES-E and fuels exhibit substantial improvements generally in all markets. Average reductions in loss function are similar for both metrics and from 10\% to 20\% in Germany and Spain and from 1\% to 5\% for Denmark and Italy. Forecast gains increase in the peak hours, as shown in columns Avg$_{8-20}$. When focusing on each individual hour, BVARXs statistically outperform VAR models and they are included in the model confidence set in most of the cases, and almost always for late morning, afternoon and evening hours. VARX models also perform accurately, but give some economically smaller gain than do BVARX models.

%
%, but interestingly with different dynamics: on one hand, improvements are found during over hours 8-24 in Germany, Denmark and Italy, whereas in Spain these refer to all hours. \textbf{This may be explained by the different generation mix XXXX and by regulatory uncertainty characterizing Italy... XXXX CHECHEECHECH}.

Going into details and exploring the forecasting ability of several models with different combinations of variables to inspect their individual contribution, we first find evidence of forecasting improvements when demand and all RES are included. %(especially during peak hours for the contribution of solar power).
Moreover, the BVAR model with only forecasted wind (besides forecasted demand and fuels) leads to better forecasts than are those obtained with the inclusion of only forecasted solar (besides forecasted demand and fuels), especially for point forecasts over hours 8-24 in Germany, Italy, and Spain (not performed in Denmark because there is no available solar power). %This highlights the interesting fact of forecasted wind having more forecasting ability than solar at peak hours. \textbf{PERCHE'?}
Comparing the ability of the BVAR model with forecasted demand and RES with the one containing forecasted demand and fuels, the former is found to perform better. However, there are further gains when all these exogenous regressors are considered simultaneously.\footnote{This may be due to the contribution of individual fuels. For instance, in an additional analysis within the frequentist approach, we have observed that models with selected fuels, forecasted demand and RES show slight improvements in the RMSEs: in Germany, the inclusion of both CO$_2$ and coal improves the forecast accuracy over hours 8-24; in Denmark, the inclusion of only gas improves the forecast accuracy during hours 8-12, whereas coal improves over the remaining ones 13-24. In Italy, coal and gas together are important during rump-up and rump-down hours (9-10 \& 18-19), whereas only gas is important during hours 11-17; this is consistent with Italy's dependence on thermal generation (and so on traditional fuels), given the still marginal penetration of RES (compared to the other countries studied). In Spain, the inclusion of coal slightly and generally improves the forecast accuracy, which is, however, comparable with the model with all fuels at selected hours (12-15 \& 23-24). In all these cases, adding the omitted fuels induces only very small reduction in the performances, hence supporting the conclusion of an overall importance of all fossil fuels when forecasting day-ahead electricity prices. These results are omitted for lack of space, but they are available on request.}
\section{Conclusions}\label{sec_Conclusions}
This paper compares the forecasting performances of linear univariate and multivariate models with enlarged specifications. Our set of models includes autoregression and vector autoregression models with only dummy variables for seasonality, which are used as baseline for the corresponding formulations enlarged by including also fuels, demand and renewable energy sources, analysed from both the frequentist and the Bayesian perspective.

%Our analysis of point and density forecasting performances cover all $24$ hours from 2015 to 2016.
Our results indicate that models with demand, renewable energy, and fuels dominate those without fuels and renewable energy sources (RES), in terms of both point and density forecasting. In particular, the first important finding is that the multivariate models outperform the univariate ones, given that they allow for interrelationships among different hours of the day. Secondly, the
Bayesian approach leads to further forecasting improvements. % in the univariate but also (and especially) in the multivariate models.%; with the Minnesota prior outperforming the conjugate Normal-Wishart.
Thirdly, and for the first time since the increasing RES penetration, we show that the models with only forecasted wind perform better than those with solar power only. And, their simultaneous inclusion further improves the performance. %Therefore, we provide an answer to the question on which RES-E forecast is more informative in which market (showing the relevance of wind over solar in all markets) and also that their inclusion does not reduce the importance of fossil fuels which are suggested to be kept in the models. \textbf{FRA: questo ultimo paragrafo potrebbe anche essere eliminato.}
%
%Additionally, we have shown that models with fossil fuels, besides forecasted demand and renewable energy sources, yield statistically more accurate point and density forecasts in all studied countries. Specifically, multivariate models which further include these exogenous drivers (that is forecasted demand and RES-E), in addition to seasonal dummies, provide forecasts statistically more accurate than those produced by the equivalent models with only monthly and weekday dummy variables during all day, consistently across the frequentist and Bayesian approach. In other words, during all hours of the day, and in particular during the peak hours, the frequentist and Bayesian VARX forecasts are more accurate than those of the corresponding frequentist and Bayesian VAR models.

We also provide a strong empirical evidence of the influence of the renewable power generation during the day, and consistently with the country intra-daily profiles. In fact, during the first hours of the day, the models without forecasted RES-E are more accurate than those with them, and again with errors from multivariate models lower than those from univariate ones. Whilst, the increasing RES-E during the day leads to more accurate forecasts from augmented models. Furthermore, our results are consistent across all adopted scoring rules, such as the RMSE and the CRPS. %These results are also confirmed from the Model Confidence Set and from the sign predictability in particular for peak hours.

%Our findings also show the importance of including RES-E and other fossil fuels when looking for the best price forecasts. We confirm that including demand forecasts and RES-E forecasts (if the market
%penetration is not negligible) leads to more accurate forecasts. More importantly, we provide an answer to the question on which RES-E forecast is more informative in which market (showing the relevance of wind over solar in all markets) and also that their inclusion does not reduce the importance of fossil fuels which are suggested to be kept in the models. \textbf{DA QUI IN SU: VERIFICARE eventuali RIPETIZIONI}

From an energy forecasting perspective these linear multivariate autoregressive models with RES, demand and fuels seem to have interesting and important advantages over the widely used univariate ones. %Encouraged by these results, our plans for the future are to include non linearities in both univariate and multivariate cases.
It is worth emphasizing the increasing relevance of density forecasting since in these recent years market operators are exploring opportunistic bidding across market sessions, as emphasized by \cite{Bunnetal2018}.
Indeed, forecasting the day-ahead prices is important for market operators and traders to plan their strategy. For example, arbitrage opportunities can be explored by deciding on which market session to bid according to the forecasted day-ahead prices. For this reason, energy regulatory authorities are trying to formulate optimal pricing rules to avoid these market inefficiencies. Agents operating balancing responsible units are exposed to economic consequences from differentials between day-ahead and balancing prices, which are used to evaluate the actual unit imbalance according to the sign of the system imbalance. In simple words, if one unit is short-imbalanced when the market is long (or long-imbalanced when the market is short), it receives profits for relieving the system (which are computed on the basis of price differentials). Otherwise, if unit and system have signs agreement, the unit receives penalties because it aggravates the system imbalance.

All these considerations clearly show the extreme relevance of both point and density forecasting for these day-ahead electricity prices and our results highlight that the Bayesian multivariate models with considered drivers improve them substantially.

\section*{Acknowledgements}
Authors thank the co-editor, seminar and conference participants at Ca' Foscari University of Venice, the EU Joint Research Centre in Ispra, the University of Warwick, the `26th Annual Symposium of the Society for Nonlinear Dynamics and Econometrics' in Tokyo, the `15th Conference on Computational Management Science' in Trondheim, the `12th Annual RCEA Bayesian Econometric Workshop' in Rimini, and the `8th Energy Finance Christmas Workshops' in Bozen for useful comments and suggestions.
This research used the SCSCF multiprocessor cluster system at Ca' Foscari University of Venice. Europe Energy S.p.A. is acknowledged for funding this research project. In addition, Angelica Gianfreda wishes to acknowledge the RTDcall2017 support for the project on \emph{Forecasting and Monitoring electricity Prices, volumes and market Mechanisms}, funded by the Free University of Bozen-Bolzano. Luca Rossini acknowledges the financial support from the EU Horizon 2020 under the Marie Sklodowska-Curie scheme (grant agreement No 796902).

%
%%%%%%%%%%%%%%%%%%%%%%%%%%%%%%%%%%%%%%%%%%%%%%%%%%%%%%
%%         		  	Bibliography	   				  %%%
%%%%%%%%%%%%%%%%%%%%%%%%%%%%%%%%%%%%%%%%%%%%%%%%%%%%%
\bibliographystyle{apalike}
\bibliography{Bib_IJF}

\newpage
%
%%%%%%%%%%%%%%%%%%%%%%%%%%%%%%%%%%%%%%%%%%%%%%%%%%%%%%
%%         		  	Appendices	   				  %%%
%%%%%%%%%%%%%%%%%%%%%%%%%%%%%%%%%%%%%%%%%%%%%%%%%%%%%
%
\newgeometry{top=.7in, bottom=.7in, left=0.7in, right=0.7in}

%\appendix{}
% APPENDIX B ON RMSES..................................
%\section{High Summary Tables of Forecasting Performances across Markets}
%\section{The Root Mean Square Errors - RMSEs}
%
\begin{table}[h!]
\centering
\caption{{\small{RMSE values for AR(VAR) benchmark models, RMSE ratios for other models}}}
%\resizebox{0.99\textwidth}{!}{
\begin{adjustbox}{width=1\textwidth,center=\textwidth}
\setlength{\tabcolsep}{0.95pt}
\begin{threeparttable}
\begin{tabular}{llllllllllcc}
\hline \\[-0.4cm]
%\multicolumn{14}{c}{Germany} \\[0.001cm]
Hour & 1 & 4 & 7 & 10 & 13 & 16 & 19 & 22 & Avg & $\mbox{Avg}_{8-20}$\\[0.01mm]
\hline  \\[-0.4cm]
\multicolumn{1}{c}{\textit{\textbf{Germany}}}\\% & & & & & & & & & & & && \\
AR & 7.240 & 7.387 & 8.027 & 8.905 & 9.214 & 9.669 & 8.692 & 6.277 & 8.259 & 9.333 \\
ARX (FD+RES+Fuels) & 5.336$^{\ast \ast \ast}$ & 5.939$^{\ast \ast}$ & \cellcolor[gray]{0.95}6.430$^{\ast \ast \ast}$ & 6.928$^{\ast \ast \ast}$ & 6.662$^{\ast \ast \ast}$ & \cellcolor[gray]{0.95}7.068$^{\ast \ast \ast}$ & \cellcolor[gray]{0.95}6.867$^{\ast \ast \ast}$ & \cellcolor[gray]{0.95}4.871$^{\ast \ast \ast}$ & 6.326 & 7.065 \\
BAR & 7.226$^{\ast \ast}$ & 7.387 & 8.011$^{\ast \ast}$ & 8.887$^{\ast \ast \ast}$ & 9.214 & 9.659$^{\ast \ast \ast}$ & 8.666$^{\ast \ast \ast}$ & 6.258$^{\ast \ast \ast}$ & 8.251 & 9.314 \\
BARX (FD+RES+Fuels) & 5.329$^{\ast \ast \ast}$ & 5.932$^{\ast \ast}$ & \cellcolor[gray]{0.95}6.430$^{\ast \ast \ast}$ & 6.768$^{\ast \ast \ast}$ & 6.597$^{\ast \ast \ast}$ & \cellcolor[gray]{0.95}7.068$^{\ast \ast \ast}$ & \cellcolor[gray]{0.95}6.728$^{\ast \ast \ast}$ & \cellcolor[gray]{0.95}4.821$^{\ast \ast \ast}$ & 6.260 & 6.972 \\
VAR & \cellcolor[gray]{0.95}4.278 & \cellcolor[gray]{0.95}4.944 & \cellcolor[gray]{0.95}6.271 & 6.905 & 7.934 & 8.350 & 8.290 & 6.164 & 6.839 & 7.993 \\
VARX (FD+RES+Fuels) & 4.492 & 5.404 & \cellcolor[gray]{0.95}6.396 & 6.083$^{\ast \ast \ast}$ & 6.315$^{\ast \ast \ast}$ & \cellcolor[gray]{0.95}7.039$^{\ast \ast \ast}$ & \cellcolor[gray]{0.95}6.715$^{\ast \ast \ast}$ & \cellcolor[gray]{0.95}4.654$^{\ast \ast \ast}$ & 5.964 & 6.698 \\
BVAR & 4.282 & \cellcolor[gray]{0.95}4.939 & \cellcolor[gray]{0.95}6.271 & 6.912 & 7.934 & 8.342 & 8.290 & 6.158 & 6.839 & 7.993 \\
BVARX (FD+RES+Fuels) & 4.445 & 5.414 & \cellcolor[gray]{0.95}6.415 & \cellcolor[gray]{0.95}6.035$^{\ast \ast \ast}$ & \cellcolor[gray]{0.95}6.276$^{\ast \ast \ast}$ & \cellcolor[gray]{0.95}6.964$^{\ast \ast \ast}$ & \cellcolor[gray]{0.95}6.591$^{\ast \ast \ast}$ & \cellcolor[gray]{0.95}4.629$^{\ast \ast \ast}$ &  5.923 & 6.642 \\
\hline \\[-0.4cm]
\multicolumn{1}{c}{\textit{\textbf{Denmark}}}\\% & & & & & & & & & & & && \\
AR & 5.850 & 6.566 & 6.857 & 13.162 & 8.226 & 8.029 & 10.300 & \cellcolor[gray]{0.95}6.012 & 8.468 & 10.465 \\
ARX (FD+RES+Fuels) & 5.376 & 6.211 & 6.082$^{\ast \ast \ast}$ & \cellcolor[gray]{0.95}9.661$^{\ast \ast \ast}$ &\cellcolor[gray]{0.95}6.893$^{\ast \ast \ast}$ & 6.544$^{\ast \ast \ast}$ & \cellcolor[gray]{0.95}8.240$^{\ast \ast \ast}$ & \cellcolor[gray]{0.95}5.345$^{\ast}$ & 6.969 & 8.121 \\
BAR & 5.838$^{\ast \ast}$ & 6.553$^{\ast \ast \ast}$ & 6.843$^{\ast \ast \ast}$ & 13.096$^{\ast \ast \ast}$ & 8.210$^{\ast \ast \ast}$ &  8.021$^{\ast \ast \ast}$ & 10.279$^{\ast \ast \ast}$ & \cellcolor[gray]{0.95}6.000$^{\ast \ast \ast}$ & 8.451 & 10.444 \\
BARX (FD+RES+Fuels) & 5.382 & 6.211 & 6.089$^{\ast \ast \ast}$ & \cellcolor[gray]{0.95}9.635$^{\ast \ast \ast}$ & \cellcolor[gray]{0.95}6.885$^{\ast \ast \ast}$ & 6.552$^{\ast \ast \ast}$ & \cellcolor[gray]{0.95}8.219$^{\ast \ast \ast}$ & \cellcolor[gray]{0.95}5.339$^{\ast}$ & 6.961 & 8.110 \\
VAR & \cellcolor[gray]{0.95}3.413 & \cellcolor[gray]{0.95}4.131 & \cellcolor[gray]{0.95}5.159 & \cellcolor[gray]{0.95}10.913 & \cellcolor[gray]{0.95}7.607 & 7.466 & 10.441 & \cellcolor[gray]{0.95}5.897 & 7.197 & 9.328 \\
VARX (FD+RES+Fuels) & 4.386 & 5.684 & 5.690 & \cellcolor[gray]{0.95}10.553 & \cellcolor[gray]{0.95}6.778$^{\ast \ast}$ & \cellcolor[gray]{0.95}6.421$^{\ast \ast \ast}$ &  8.750$^{\ast \ast \ast}$ & \cellcolor[gray]{0.95}5.213$^{\ast \ast \ast}$ & 6.974 & 8.460 \\
BVAR & \cellcolor[gray]{0.95}3.413 & \cellcolor[gray]{0.95}4.135& \cellcolor[gray]{0.95}5.164 & \cellcolor[gray]{0.95}10.913 & \cellcolor[gray]{0.95}7.599$^{\ast \ast \ast}$ &7.451$^{\ast \ast \ast}$ & 10.431$^{\ast \ast}$ & \cellcolor[gray]{0.95}5.891$^{\ast \ast}$ & 7.190 & 9.319 \\
BVARX (FD+RES+Fuels) &4.386 & 5.680 & 5.690 & \cellcolor[gray]{0.95}10.553 & \cellcolor[gray]{0.95}6.755$^{\ast \ast \ast}$ & \cellcolor[gray]{0.95}6.421$^{\ast \ast \ast}$ & \cellcolor[gray]{0.95}8.760$^{\ast \ast \ast}$ & \cellcolor[gray]{0.95}5.201$^{\ast \ast \ast}$ & 6.974 &  8.451 \\
\hline \\[-0.4cm]
\multicolumn{1}{c}{\textit{\textbf{Italy}}}\\% & & & & & & & & & & & && \\
AR & 4.560 & \cellcolor[gray]{0.95}4.405 & \cellcolor[gray]{0.95}5.162 & \cellcolor[gray]{0.95}9.107 & 6.389 & 8.122 & \cellcolor[gray]{0.95}9.835 & \cellcolor[gray]{0.95}7.104 & 6.838 & 8.331 \\
ARX (FD+RES+Fuels) & 4.410 & \cellcolor[gray]{0.95}4.185 & \cellcolor[gray]{0.95}4.966 & \cellcolor[gray]{0.95}8.561 & 5.776$^{\ast \ast \ast}$ & \cellcolor[gray]{0.95}7.513$^{\ast \ast}$ & \cellcolor[gray]{0.95}9.638 & \cellcolor[gray]{0.95}6.862 & 6.469 & 7.806 \\
BAR &  4.551 & \cellcolor[gray]{0.95}4.401 & \cellcolor[gray]{0.95}5.157 & \cellcolor[gray]{0.95}9.098$^{\ast \ast}$ & 6.389 & 8.122$^{\ast}$ & \cellcolor[gray]{0.95}9.835 & \cellcolor[gray]{0.95}7.111 &  6.831 & 8.331 \\
BARX (FD+RES+Fuels) &4.432 & \cellcolor[gray]{0.95}4.238 & \cellcolor[gray]{0.95}4.961$^{\ast}$ & \cellcolor[gray]{0.95}8.524$^{\ast \ast}$ & \cellcolor[gray]{0.95}5.756$^{\ast \ast \ast}$ & \cellcolor[gray]{0.95}7.488$^{\ast \ast \ast}$ & \cellcolor[gray]{0.95}9.560$^{\ast}$ & \cellcolor[gray]{0.95}6.870 & 6.455 &  7.764 \\
VAR & \cellcolor[gray]{0.95}3.884 & \cellcolor[gray]{0.95}4.105 & \cellcolor[gray]{0.95}4.753 & \cellcolor[gray]{0.95}8.569 & \cellcolor[gray]{0.95}6.144 & \cellcolor[gray]{0.95}7.650 & \cellcolor[gray]{0.95}9.582 & \cellcolor[gray]{0.95}6.893 & 6.507 & 7.901 \\
VARX (FD+RES+Fuels) & 4.094 & \cellcolor[gray]{0.95}4.228 & \cellcolor[gray]{0.95}4.881 & \cellcolor[gray]{0.95}8.098 & \cellcolor[gray]{0.95}5.511$^{\ast \ast \ast}$ & \cellcolor[gray]{0.95}7.107$^{\ast \ast}$ & 9.668 & 6.996 & 6.357 & 7.530 \\
BVAR & \cellcolor[gray]{0.95}3.880 & \cellcolor[gray]{0.95}4.101 & \cellcolor[gray]{0.95}4.753 & \cellcolor[gray]{0.95}8.569 & 6.150 & \cellcolor[gray]{0.95}7.650 & \cellcolor[gray]{0.95}9.572$^{\ast \ast}$ & \cellcolor[gray]{0.95}6.893$^{\ast}$ & 6.507 & 7.901 \\
BVARX (FD+RES+Fuels) & 4.051 & \cellcolor[gray]{0.95}4.142 & \cellcolor[gray]{0.95}4.853 & \cellcolor[gray]{0.95}8.106 & \cellcolor[gray]{0.95}5.523$^{\ast \ast \ast}$ & \cellcolor[gray]{0.95}7.061$^{\ast \ast}$ & 9.649 & 6.934 & 6.325 & 7.506 \\
\hline \\[-0.4cm]
\multicolumn{1}{c}{\textit{\textbf{Spain}}}\\% & & & & & & & & & & & && \\
AR & 7.036 & 6.741 & 7.066 & 6.421 & 6.140 & 6.873 & 5.574 & \cellcolor[gray]{0.95}4.628 & 6.299 & 6.389 \\
ARX (FD+RES+Fuels) & 0.809$^{\ast \ast \ast}$ & 0.779$^{\ast \ast \ast}$ & 0.776$^{\ast \ast \ast}$ & 0.804$^{\ast \ast \ast}$ & 0.813$^{\ast \ast \ast}$ & 0.796$^{\ast \ast \ast}$ & \cellcolor[gray]{0.95}0.883$^{\ast \ast}$ & \cellcolor[gray]{0.95}1.020 & 0.825 & 0.814 \\
BAR & 0.999$^{\ast}$ & 0.998$^{\ast \ast \ast}$ & 0.997$^{\ast \ast \ast}$ & 0.994$^{\ast \ast \ast}$ & 0.997$^{\ast \ast \ast}$ & 0.997$^{\ast \ast \ast}$ & 0.994$^{\ast \ast \ast}$ & \cellcolor[gray]{0.95}0.996$^{\ast \ast \ast}$ & 0.996 & 0.996 \\
BARX (FD+RES+Fuels) & 0.818$^{\ast \ast \ast}$ & 0.798$^{\ast \ast \ast}$ & 0.805$^{\ast \ast \ast}$ & 0.844$^{\ast \ast \ast}$ & 0.838$^{\ast \ast \ast}$ & 0.815$^{\ast \ast \ast}$ & \cellcolor[gray]{0.95}0.903$^{\ast \ast}$ & \cellcolor[gray]{0.95}1.040 & 0.847 & 0.839 \\
VAR & \cellcolor[gray]{0.95}3.943 & \cellcolor[gray]{0.95}4.638 & 5.227 & 4.761 & 5.018 & 5.908 & 5.363 & \cellcolor[gray]{0.95}4.823 & 5.110 & 5.317 \\
VARX (FD+RES+Fuels) & \cellcolor[gray]{0.95}1.001 & \cellcolor[gray]{0.95}0.940 & \cellcolor[gray]{0.95}0.843$^{\ast \ast \ast}$ & \cellcolor[gray]{0.95}0.855$^{\ast \ast \ast}$ & \cellcolor[gray]{0.95}0.825$^{\ast \ast \ast}$ & \cellcolor[gray]{0.95}0.752$^{\ast \ast \ast}$ & \cellcolor[gray]{0.95}0.791$^{\ast \ast \ast}$ & \cellcolor[gray]{0.95}0.858$^{\ast \ast}$ & 0.834 & 0.801 \\
BVAR & \cellcolor[gray]{0.95}1.000 & \cellcolor[gray]{0.95}1.000 & 1.000 & 0.998 & 0.999 & 0.997$^{\ast \ast}$ & 0.993$^{\ast \ast \ast}$ & \cellcolor[gray]{0.95}0.994$^{\ast \ast \ast}$ & 0.998 & 0.997 \\
BVARX (FD+RES+Fuels) & \cellcolor[gray]{0.95}1.010 & \cellcolor[gray]{0.95}0.952$^{\ast}$ & 0.856$^{\ast \ast \ast}$ & \cellcolor[gray]{0.95}0.860$^{\ast \ast \ast}$ & \cellcolor[gray]{0.95}0.829$^{\ast \ast \ast}$ & \cellcolor[gray]{0.95}0.757$^{\ast \ast \ast}$ & \cellcolor[gray]{0.95}0.797$^{\ast \ast \ast}$ & \cellcolor[gray]{0.95}0.860$^{\ast \ast}$ & 0.841 & 0.806 \\
\hline
\end{tabular}
\small{
\begin{tablenotes}
\item \textit{Notes:}
\item[1] Forecast errors are calculated using rolling window estimation. `Avg' and `Avg$_{8-20}$' stand for RMSEs computed as in \eqref{RMSE_Avg} and \eqref{RMSE_Avg_Red}.
 \item[2] Please refer to Section \ref{sec_Models} for details on model formulations. The `X' indicates models with exogenous variables, while `B' Bayesian conjugate Normal-Wishart priors.
 %\item[3] For (V)AR baseline models, table reports RMSEs; for all other (V)ARX models, table reports ratio between RMSE of current model and RMSE of (V)AR benchmark. Entries less than 1 indicate that forecasts from current model are more accurate than forecasts from corresponding baseline model.
\item[3] $^{\ast \ast \ast}$, $^{\ast \ast}$ and $^{\ast}$ indicate RMSE ratios are significantly different from 1 at $1\%$, $5\%$ and $10\%$, according to Diebold-Mariano test.
\item[4] Gray cells indicate those models that belong to the Superior Set of Models delivered by the Model Confidence Set procedure at confidence level $10\%$.%\sout{To provide a rough gauge of whether the RMSE ratios are significantly different from 1, we use the Diebold-Mariano t-statistic for equal RMSE. Differences in accuracy that are statistically different from zero are denoted by asterisks, corresponding to significance levels of $^{\ast} 10\%$, $^{\ast \ast} 5\%$ and $^{\ast \ast \ast} 1\%$. The underlying p-values are based on t-statistics computed with a serial correlation-robust variance, using the pre-whitened quadratic spectral estimator of Andrews and Monahan (1992). For all the models, since they are nested, we report $p$-values based on one-sided tests, takin the AR (VAR) as the null and the other model in question as alternative}
\end{tablenotes}
}
\end{threeparttable}
%      }
\end{adjustbox}
\label{table:RMSE_Rol}
\end{table}

\clearpage

\begin{table}[h!]
\centering
\caption{{\small{Average CRPS for AR(VAR) benchmark model, CRPS ratios for other models.}}}
%\resizebox{0.99\textwidth}{!}{
\begin{adjustbox}{width=1\textwidth,center=\textwidth}
\setlength{\tabcolsep}{0.95pt}
\begin{threeparttable}
\begin{tabular}{llllllllllcc}
\hline \\[-0.4cm]
%\multicolumn{14}{c}{Germany} \\[0.001cm]
Hour & 1 & 4 & 7 & 10 & 13 & 16 & 19 & 22 & Avg & $\mbox{Avg}_{8-20}$\\[0.01mm]
\hline  \\[-0.4cm]
\multicolumn{1}{c}{\textit{\textbf{Germany}}}\\% & & & & & & & & & & & && \\
AR & 3.770 & 4.062 & 4.467 & 4.942 & 4.962 & 4.970 & 4.792 & 3.423 & 4.427 & 4.964 \\
ARX (FD+RES+Fuels) & 2.926$^{\ast \ast \ast}$ & 3.420$^{\ast \ast \ast}$ & 3.672$^{\ast \ast \ast}$ & 3.781$^{\ast \ast \ast}$ & 3.563$^{\ast \ast \ast}$ & \cellcolor[gray]{0.95}3.593$^{\ast \ast \ast}$ & \cellcolor[gray]{0.95}3.786$^{\ast \ast \ast}$ & \cellcolor[gray]{0.95}2.663$^{\ast \ast \ast}$ & 3.422 & 3.733 \\
BAR & 3.770 & 4.062 & 4.458$^{\ast}$ & 4.927$^{\ast \ast \ast}$ & 4.957$^{\ast}$ & 4.960$^{\ast \ast}$ & 4.768$^{\ast \ast \ast}$ & 3.409$^{\ast \ast \ast}$ & 4.418 & 4.954 \\
BARX (FD+RES+Fuels) &  2.926$^{\ast \ast \ast}$ & 3.420$^{\ast \ast \ast}$ & 3.672$^{\ast \ast \ast}$ & 3.702$^{\ast \ast \ast}$ & 3.528$^{\ast \ast \ast}$ & \cellcolor[gray]{0.95}3.598$^{\ast \ast \ast}$ & \cellcolor[gray]{0.95}3.719$^{\ast \ast \ast}$ & \cellcolor[gray]{0.95}2.646$^{\ast \ast \ast}$ & 3.391 & 3.688 \\
VAR & 2.261 & 2.901 & \cellcolor[gray]{0.95}3.443 & 3.772 & 4.185 & 4.208 & 4.525 & 3.347 & 3.643 & 4.173 \\
VARX (FD+RES+Fuels) & 2.390 & 3.040 & \cellcolor[gray]{0.95}3.443 & 3.316$^{\ast \ast \ast}$ & 3.294$^{\ast \ast \ast}$ & \cellcolor[gray]{0.95}3.509$^{\ast \ast \ast}$ & \cellcolor[gray]{0.95}3.692$^{\ast \ast \ast}$ & \cellcolor[gray]{0.95}2.570$^{\ast \ast \ast}$ & 3.169 & 3.497 \\
BVAR & \cellcolor[gray]{0.95}2.254$^{\ast \ast}$ & \cellcolor[gray]{0.95}2.886$^{\ast \ast \ast}$ & \cellcolor[gray]{0.95}3.426$^{\ast \ast \ast}$ & 3.764$^{\ast \ast}$ & 4.177$^{\ast \ast}$ & 4.191$^{\ast \ast \ast}$ & 4.507$^{\ast \ast \ast}$ & 3.334$^{\ast \ast \ast}$ & 3.632 & 4.160 \\
BVARX (FD+RES+Fuels) & 2.358 & 3.037 & \cellcolor[gray]{0.95}3.426 & \cellcolor[gray]{0.95}3.282$^{\ast \ast \ast}$ & \cellcolor[gray]{0.95}3.260$^{\ast \ast \ast}$ & \cellcolor[gray]{0.95}3.472$^{\ast \ast \ast}$ & \cellcolor[gray]{0.95}3.615$^{\ast \ast \ast}$ & \cellcolor[gray]{0.95}2.554$^{\ast \ast \ast}$ & 3.137 & 3.451 \\
\hline \\[-0.4cm]
\multicolumn{1}{c}{\textit{\textbf{Denmark}}}\\% & & & & & & & & & & & && \\
AR & 3.236 & 3.690 & 3.896 & 8.844 & 4.400 & 4.156 & 5.379 & 3.188 & 4.901 & 6.199 \\
ARX (FD+RES+Fuels) & 2.997$^{\ast \ast \ast}$ & 3.446$^{\ast \ast \ast}$ & 3.471$^{\ast \ast \ast}$ & \cellcolor[gray]{0.95}7.517$^{\ast \ast \ast}$ & 3.670$^{\ast \ast \ast}$ & \cellcolor[gray]{0.95}3.387$^{\ast \ast \ast}$ & \cellcolor[gray]{0.95}4.255$^{\ast \ast \ast}$ & 2.866$^{\ast \ast \ast}$ & 4.210 &  5.151 \\
BAR & 3.230$^{\ast \ast}$ & 3.683$^{\ast \ast}$ & 3.892 & 8.817$^{\ast \ast \ast}$ & 4.387$^{\ast \ast \ast}$ & 4.148$^{\ast \ast}$ & 5.368$^{\ast \ast}$ & 3.182$^{\ast \ast}$ & 4.891 & 6.187 \\
BARX (FD+RES+Fuels) & 2.997$^{\ast \ast \ast}$ & 3.443$^{\ast \ast \ast}$ & 3.475$^{\ast \ast \ast}$ & \cellcolor[gray]{0.95}7.509$^{\ast \ast \ast}$ & 3.665$^{\ast \ast \ast}$ & 3.387$^{\ast \ast \ast}$ & \cellcolor[gray]{0.95}4.249$^{\ast \ast \ast}$ & 2.866$^{\ast \ast \ast}$ & 4.210 & 5.151 \\
VAR & 2.019 & 2.644 & 3.035 & \cellcolor[gray]{0.95}7.829 & 3.925 & 3.761 & 5.329 & 3.100 & 4.273 & 5.606 \\
VARX (FD+RES+Fuels) & 2.314 & 3.093 & 3.199 & \cellcolor[gray]{0.95}7.727$^{\ast}$ & 3.556$^{\ast \ast \ast}$ & \cellcolor[gray]{0.95}3.295$^{\ast \ast \ast}$ & 4.487$^{\ast \ast \ast}$ & 2.737$^{\ast \ast \ast}$ & 4.119 & 5.219 \\
BVAR & \cellcolor[gray]{0.95}2.007$^{\ast \ast \ast}$ & \cellcolor[gray]{0.95}2.631$^{\ast \ast \ast}$ & \cellcolor[gray]{0.95}3.020$^{\ast \ast \ast}$ & \cellcolor[gray]{0.95}7.790$^{\ast \ast \ast}$ & 3.901$^{\ast \ast \ast}$ & 3.731$^{\ast \ast \ast}$ & 5.308$^{\ast \ast \ast}$ & 3.088$^{\ast \ast \ast}$ & 4.247 & 5.572 \\
BVARX (FD+RES+Fuels) & 2.300 & 3.067 & 3.178 & \cellcolor[gray]{0.95}7.743$^{\ast}$ & \cellcolor[gray]{0.95}3.532$^{\ast \ast \ast}$ & \cellcolor[gray]{0.95}3.268$^{\ast \ast \ast}$ & 4.460$^{\ast \ast \ast}$ & \cellcolor[gray]{0.95}2.725$^{\ast \ast \ast}$ & 4.106 & 5.208 \\
\hline \\[-0.4cm]
\multicolumn{1}{c}{\textit{\textbf{Italy}}}\\% & & & & & & & & & & & && \\
AR & 2.547 & 2.460 & 2.851 & 4.772 & 3.494 & 4.390 & \cellcolor[gray]{0.95}5.037 & \cellcolor[gray]{0.95}3.616 & 3.657 & 4.413 \\
ARX (FD+RES+Fuels) & 2.476 & \cellcolor[gray]{0.95}2.362$^{\ast}$ & \cellcolor[gray]{0.95}2.731$^{\ast \ast}$ & \cellcolor[gray]{0.95}4.514$^{\ast \ast}$ & \cellcolor[gray]{0.95}3.141$^{\ast \ast \ast}$ & \cellcolor[gray]{0.95}4.100$^{\ast \ast \ast}$ & \cellcolor[gray]{0.95}4.992 & \cellcolor[gray]{0.95}3.522$^{\ast \ast}$ & 3.481 & 4.157 \\
BAR & 2.539$^{\ast \ast}$ & 2.455 & 2.854 & 4.767 & 3.491 & 4.386 & \cellcolor[gray]{0.95}5.032 & \cellcolor[gray]{0.95}3.609$^{\ast \ast}$ & 3.653 & 4.409 \\
BARX (FD+RES+Fuels) & 2.488 & \cellcolor[gray]{0.95}2.386 & \cellcolor[gray]{0.95}2.720$^{\ast \ast}$ & \cellcolor[gray]{0.95}4.457$^{\ast \ast \ast}$ & \cellcolor[gray]{0.95}3.113$^{\ast \ast \ast}$ & \cellcolor[gray]{0.95}4.039$^{\ast \ast \ast}$ & \cellcolor[gray]{0.95}4.881$^{\ast \ast}$ & \cellcolor[gray]{0.95}3.511$^{\ast \ast}$ & 3.449 & 4.095 \\
VAR & \cellcolor[gray]{0.95}2.147 & \cellcolor[gray]{0.95}2.265 & \cellcolor[gray]{0.95}2.588 & \cellcolor[gray]{0.95}4.509 & 3.342 & \cellcolor[gray]{0.95}4.095 & \cellcolor[gray]{0.95}4.954 & \cellcolor[gray]{0.95}3.569 & 3.469 & 4.177 \\
VARX (FD+RES+Fuels) & 2.291 & \cellcolor[gray]{0.95}2.378 & \cellcolor[gray]{0.95}2.694 & \cellcolor[gray]{0.95}4.356 & \cellcolor[gray]{0.95}3.008$^{\ast \ast \ast}$ & \cellcolor[gray]{0.95}3.812$^{\ast \ast}$ & \cellcolor[gray]{0.95}5.177 & \cellcolor[gray]{0.95}3.744 & 3.455 & 4.043 \\
BVAR & \cellcolor[gray]{0.95}2.145 & \cellcolor[gray]{0.95}2.263 & \cellcolor[gray]{0.95}2.580$^{\ast}$ & \cellcolor[gray]{0.95}4.482$^{\ast \ast \ast}$ & 3.329$^{\ast \ast}$ & \cellcolor[gray]{0.95}4.075$^{\ast \ast \ast}$ & \cellcolor[gray]{0.95}4.924$^{\ast \ast \ast}$ & \cellcolor[gray]{0.95}3.551$^{\ast \ast \ast}$ & 3.452 & 4.156 \\
BVARX (FD+RES+Fuels) & 2.259 & \cellcolor[gray]{0.95}2.310 & \cellcolor[gray]{0.95}2.653 & \cellcolor[gray]{0.95}4.338 & \cellcolor[gray]{0.95}2.991$^{\ast \ast \ast}$ & \cellcolor[gray]{0.95}3.763$^{\ast \ast \ast}$ & \cellcolor[gray]{0.95}5.157 & \cellcolor[gray]{0.95}3.708 & 3.417 & 4.014 \\
\hline \\[-0.4cm]
\multicolumn{1}{c}{\textit{\textbf{Spain}}}\\% & & & & & & & & & & & && \\
AR & 3.914 & 3.757 & 3.948 & 3.555 & 3.402 & 3.844 & 3.139 & \cellcolor[gray]{0.95}2.669 & 3.517 & 3.556 \\
ARX (FD+RES+Fuels) & 3.112$^{\ast \ast \ast}$ & 2.915$^{\ast \ast \ast}$ & 3.064$^{\ast \ast \ast}$ & 2.901$^{\ast \ast \ast}$ & 2.786$^{\ast \ast \ast}$ & 3.037$^{\ast \ast \ast}$ & \cellcolor[gray]{0.95}2.734$^{\ast \ast \ast}$ & \cellcolor[gray]{0.95}2.613 & 2.891 & 2.895 \\
BAR & 3.910 & 3.746 & 3.952 & 3.541 & 3.388 & 3.832 & 3.117 & \cellcolor[gray]{0.95}2.658 & 3.503 & 3.538 \\
BARX (FD+RES+Fuels) & 3.151$^{\ast \ast \ast}$ & 2.987$^{\ast \ast \ast}$ & 3.190$^{\ast \ast \ast}$ & 3.040$^{\ast \ast \ast}$ & 2.858$^{\ast \ast \ast}$ & 3.106$^{\ast \ast \ast}$ & 2.787$^{\ast \ast \ast}$ & \cellcolor[gray]{0.95}2.666 & 2.961 & 2.973 \\
VAR & \cellcolor[gray]{0.95}2.128 & 2.536 & 2.897 & 2.644 & 2.743 & 3.267 & 3.019 & \cellcolor[gray]{0.95}2.745 & 2.831 & 2.945 \\
VARX (FD+RES+Fuels) & \cellcolor[gray]{0.95}2.145 & \cellcolor[gray]{0.95}2.381$^{\ast}$ & \cellcolor[gray]{0.95}2.422$^{\ast \ast \ast}$ & \cellcolor[gray]{0.95}2.239$^{\ast \ast \ast}$ & \cellcolor[gray]{0.95}2.271$^{\ast \ast \ast}$ & \cellcolor[gray]{0.95}2.447$^{\ast \ast \ast}$ & \cellcolor[gray]{0.95}2.364$^{\ast \ast \ast}$ & \cellcolor[gray]{0.95}2.306$^{\ast \ast \ast}$ & 2.350 & 2.350 \\
BVAR & \cellcolor[gray]{0.95}2.122 & \cellcolor[gray]{0.95}2.526 & 2.885 & 2.631 & 2.724 & 3.244 & 2.989 & \cellcolor[gray]{0.95}2.720 & 2.814 & 2.924 \\
BVARX (FD+RES+Fuels) & \cellcolor[gray]{0.95}2.164 & \cellcolor[gray]{0.95}2.414 & 2.471$^{\ast \ast \ast}$ & \cellcolor[gray]{0.95}2.250$^{\ast \ast \ast}$ & \cellcolor[gray]{0.95}2.263$^{\ast \ast \ast}$ & \cellcolor[gray]{0.95}2.440$^{\ast \ast \ast}$ & \cellcolor[gray]{0.95}2.373$^{\ast \ast \ast}$ & \cellcolor[gray]{0.95}2.306$^{\ast \ast \ast}$ & 2.361 & 2.353 \\
\hline
\end{tabular}
\small{
\begin{tablenotes}
\item \textit{Notes:}
\item[1] Forecast errors are calculated using rolling window estimation. `Avg' and `Avg$_{8-20}$' stand for average CRPS for average values.
 \item[2] Please refer to Section \ref{sec_Models} for details on model formulations. The `X' indicates models with exogenous variables, while `B' Bayesian conjugate Normal-Wishart priors.
% \item[3] For (V)AR baseline models, table reports average CRPS; for all other (V)ARX models, table reports ratio between average CRPS of current model and average CRPS of (V)AR benchmark. Entries less than 1 indicate that forecasts from current model are more accurate than forecasts from corresponding baseline model.
\item[3] $^{\ast \ast \ast}$, $^{\ast \ast}$ and $^{\ast}$ indicate average score ratios are significantly different from 1 at $1\%$, $5\%$ and $10\%$, according to Diebold-Mariano test.
\item[4] Gray cells indicate those models that belong to the Superior Set of Models delivered by the Model Confidence Set procedure at confidence level $10\%$.
\end{tablenotes}
}
\end{threeparttable}
\end{adjustbox}
%      }
\label{table:CRPS_Rol}
\end{table}
\clearpage

\end{document}